\documentstyle[12pt]{article}
\catcode`\@=11
\def\numberbysection{\@addtoreset{equation}{section}
        \def\theequation{\thesection.\arabic{equation}}}
\def\beq{\begin{equation}}
\def\eeq{\end{equation}}
\def\barr{\begin{eqnarray}}
\def\earr{\end{eqnarray}}
\def\winf{W_{1+\infty}\ }
\def\u1{\widehat{U(1)}}
\def\su2{\widehat{SU(2)}_1}
\def\suem{\widehat{SU(m)}_1}
\def\rep{representation }
\def\reps{representations }

\def\disp{\displaystyle}

\def\I{{\rm Im}\ }
\def\R{{\rm Re}\ }
\def\Z{{\bf Z}}
\newcommand{\secn}[1]{Section~\ref{#1}}

\newcommand{\eq}[1]{Eq.~(\ref{#1})}

\newcommand{\nl}{\nonumber \\}

\renewcommand{\a}{\alpha}
\renewcommand{\b}{\beta}

\numberbysection
\begin{document}
\begin{titlepage}
\begin{center}
\hfill  \quad DFF 249/5/96 \\
\hfill  \quad hep-th/9506127 \\
\vskip .3 in
{\LARGE Modular Invariant Partition Functions}

{\LARGE in the Quantum Hall Effect }
\vskip 0.2in
Andrea CAPPELLI \\
{\em I.N.F.N. and Dipartimento di Fisica, Largo E. Fermi 2,
 I-50125 Firenze, Italy}
\\
\vskip 0.1in
Guillermo~R.~ZEMBA \\%[.2in]
{\em Comisi\'on Nacional de Energ\'{\i}a At\'omica and
Instituto Balseiro, Universidad Nacional de Cuyo,
Centro At\'omico Bariloche, 8400 - San Carlos de Bariloche,
R\'{\i}o Negro, Argentina}
\end{center}
\vskip .1 in
\begin{abstract}
We study the partition function for the low-energy edge excitations 
of the incompressible electron fluid. 
On an annular geometry, these excitations have opposite chiralities
on the two edges; thus, the partition function takes the standard form 
of rational conformal field theories. 
In particular, it is invariant under modular transformations of the toroidal 
geometry made by the angular variable and the compact Euclidean time. 
The Jain series of plateaus have been described by two types of 
edge theories: the minimal models of the $\winf$ algebra
of quantum area-preserving diffeomorphisms, and their non-minimal
version, the theories with $\u1\times\suem$ affine algebra.
We find modular invariant partition functions for the latter models.
Moreover, we relate the Wen topological order to
the modular transformations and the Verlinde fusion algebra.
We find new, non-diagonal modular invariants which describe edge theories
with extended symmetry algebra; their Hall conductivities match the 
experimental values beyond the Jain series.
\end{abstract}
\vskip 1.cm
\centerline{\it To Claude Itzykson }
\vfill
\hfill May 1996
\end{titlepage}
\pagenumbering{arabic}
%
%- 1 ------------

\section{Introduction}
\label{intro}

At the plateaus of the quantum Hall effect \cite{prange}, the electrons form 
an incompressible fluid \cite{laugh}, in which $(2+1)$-dimensional density 
waves have a high gap. At energies below this gap, there exist 
$(1+1)$-dimensional, chiral excitations at the edge of the sample, which 
can be described by a low-energy effective field theory \cite{wen}.
This is the $(1+1)$-dimensional conformal theory \cite{gins} on the edge 
\cite{cdtz1}, which is equivalent to the topological $(2+1)$-dimensional 
Chern-Simons theory \cite{juerg}. 
For example, the Laughlin plateaus with Hall conductivity
$\sigma_H=(e^2/h)\nu$ , $\nu=1,1/3,1/5,\dots$ have been successfully
described by the Abelian Chern-Simons theory, or by the
corresponding conformal field theory with $\u1$ current
algebra and Virasoro central charge $c=1$.
The spectrum predicted by these theories has been
confirmed by recent experiments \cite{tdom}\cite{tunn}\cite{milli}.

The quantum numbers of edge excitations always take rational
values; this suggests that the conformal field theories describing them
should belong to the special, well understood class of
{\it rational } conformal field theories (RCFT) \cite{mose}.
By definition, a RCFT contains a {\it finite} number of
representations of the (chiral) symmetry algebra, which is the
Virasoro algebra extended by other currents in the theory. 
These representations are encoded in the partition function 
\cite{cardy} of the Euclidean theory defined on the geometry of a 
space-time torus $S^1\times S^1$.
This gives a precise inventory of all the states in the
Hilbert space and serves as a {\it definition} of the RCFT. 
Clearly, each representation of the extended symmetry algebra
describes a sector of the Hilbert space which contains infinite states. 
In the literature, it was shown that the rational spectrum follows from 
the extended symmetry and, moreover, from the invariance 
of the partition function under modular transformations of 
the periods of the torus \cite{cardy}\cite{ciz}, which span the discrete group
$\Gamma=SL(2,\Z)/\Z_2$.
Furthermore, Verlinde \cite{verl}\cite{mose} has shown the relation between 
modular invariance and the {\it fusion rules} of the extended symmetry
algebra, which are the selection rules for making composite states.
Moreover, Witten \cite{jones}\cite{c-s} has explained that any RCFT is 
associated to a Chern-Simons theory, and that the torus partition function
in the former theory corresponds to a path-integral amplitude for
the latter theory on the manifold
$S^1\times S^1\times {\bf R}$, where ${\bf R}$ is the time axis.

In the quantum Hall effect, a complete description of the
edge excitations similarly requires the construction of their 
partition function.
In this paper, we consider a spatial annulus with Euclidean compact time:
this space-time manifold is topologically ${\cal M}=S^1\times S^1\times I$,
because the radial coordinate runs over the finite interval 
$I$ and the angular coordinate and Euclidean time are both compact.
We thus find the partition function of the conformal field theory
on the edge(s) $\partial{\cal M}=S^1\times S^1$, i.e. a space-time torus.
Due to the Witten relation, this is also a path-integral of the
Chern-Simons theory on the manifold ${\cal M}$, modulo a convenient
renaming of the radial and Euclidean time coordinates.

This annulus partition function takes the standard sesquilinear form 
\cite{cardy} in terms of the characters of the chiral and antichiral algebras,
which pertain to the inner and outer edges, respectively.
In section two, we build the partition functions for the
simpler Laughlin plateaus $\nu=1/p$, $p$ odd, starting from
raw conformal field theory data, namely the
characters of the affine Abelian algebra $\u1$ \cite{gins}.
The invariance under transformations of the modular parameter
$\tau$ of the space-time torus $S^1\times S^1$ is solved in
the standard way \cite{ciz}. 
Actually, the presence of fermionic excitations in
the quantum Hall effect implies the invariance under the
$\Gamma (2)$ subgroup of the modular group, as in the Neveu-
Schwarz sector of super-symmetric conformal theories \cite{gins}.
The charge spectrum is constrained by additional modular conditions,
which have also been found \cite{c-s} in the Chern-Simons quantization 
on ${\cal M}$. 
We find that the maximal chiral algebra of the RCFT is the extension of
$\u1$ by a current of Virasoro dimension $h=p/2$, and the
Verlinde fusion rules are the Abelian group $\Z_p$ (addition modulo
$p$). This is an {\it odd} variant of the free boson theory
compactified on a rational torus \cite{gins}.

A nice property of the annulus partition function is that it encodes
the Wen {\it topological order} \cite{topord}, which is the degeneracy 
of the quantum Hall
ground state on the (ideal) compact surface $\Sigma_g$ of genus $g$.
This degeneracy is accounted for by the effective Chern-Simons theory,
where it corresponds \cite{c-s} to the dimension of the Hilbert space of 
topologically non-trivial gauge fields on $\Sigma_g\times {\bf R}$.
This dimension can be readily computed from the RCFT fusion rules and the
$S$ modular transformation, by using the Verlinde formulas and 
the Witten relation, suitably extended to manifolds with boundaries.
We show that the Wen topological order can also be defined
for the annulus geometry, and that it always takes the same value of the 
toroidal geometry $\Sigma_1$.
Moreover, the Verlinde formulas immediately yield the 
topological order in presence of static impurities in the 
Hall sample, a quantity which would otherwise need the explicit
calculation of Chern-Simons amplitudes with Wilson lines.
For Abelian theories, we show that the topological
order is independent of the impurities, thus generalising
previous perturbative results \cite{topord}.

The Hall effect of spin-polarised electrons at the stable plateaus
$\nu=m/(ms \pm 1)$, with $m =2,3,\dots$ and $s$ even, 
is well understood in terms of the Jain trial wave functions \cite{jain}.
Two specific types of edge conformal field theories have been proposed.
The first ones are characterised by the dynamical
symmetry of the incompressible fluid under area-preserving
diffeomorphisms of the plane \cite{ctz1}\cite{sakita}, 
which implies \cite{ctz3} the quantum algebra $\winf$ \cite{kac1}.
Actually, the Jain plateaus are in one-to-one correspondence
with the {\it $\winf$ minimal models } \cite{ctz5}, which are $c=m$ 
conformal field theories made by fully degenerate $\winf$ 
representations \cite{kac2}.
The second proposal \cite{juerg}\cite{zee} is based on the multi-component 
Abelian Chern-Simons theories, which correspond to RCFTs with 
$\u1^{\otimes m}$ affine algebras \cite{wen}.
In particular, the Jain series were shown to correspond to the
special cases for which the $\u1^{\otimes m}$ algebra extends 
to the $\u1\times\suem$ one.

In \secn{sec3}, we recall some properties of these two classes of
conformal field theories.
The $\u1\times\suem$ theory is a non-minimal version of the 
$\winf$ theory, which displays the same spectrum of charge and fractional 
spin of excitations, but with different multiplicities \cite{ctz5}.
We explain that the non-minimal and minimal theories have full and
``hidden'' $SU(m)$ symmetry, respectively.
For $m=2$, we discuss in detail the reduction of degrees of
freedom connecting the former to the latter, and the different
properties of excitations in the two theories.
In the literature, the non-minimal theory was first introduced and is
more widely accepted, although the physical origin of the $SU(m)$
symmetry at the Jain plateaus is not well understood.
We remark that for $m=2$, at least, the $SU(2)$ symmetry can be associated 
to the electron spin, which characterises the excitations above the different,
spin-singlet ground state at $\nu=2/3$ \cite{spinsing}, which has been 
observed in samples of low electron density \cite{tilt}.

In \secn{sec4}, we construct the partition functions for
the $\u1\times\suem$ theories. 
The simple left-right diagonal invariant is found for any
$m=2,3,\dots$; actually, it exists for any $\u1^{\otimes m}$ theory,
as also discussed in Refs.\cite{esko}\cite{read}.
The spectrum of edge excitations of this partition function matches
those described in earlier studies, with, in addition, a
pairing of excitations between the two edges.
We easily compute the topological order from the modular transformations
and recover the result of previous explicit Chern-Simons calculations
\cite{topord}.
On the other hand, we do not find  modular invariant partition functions
corresponding to the $\winf$ minimal models, which implies that 
these are not RCFTs. 
If the Jain plateaus do not have the full $SU(m)$ symmetry, then the
minimal models are physical, and their partition functions should be found 
by a novel approach, in which modular invariance would be appropriately 
generalised or replaced by other requirements.
We will not address these issues here, besides showing that is not correct 
to simply dispose of modular invariance. 
Actually, in Appendix B we construct the modular variant partition
functions fulfilling the remaining building and self-consistency criteria. 
We find too many solutions for the values of the filling fraction, 
in disagreement with the few experimental points. 

One of the main questions which motivated this work concerns the existence 
of other {\it non-diagonal} solutions to the modular invariance conditions,
defining new RCFTs, which would describe further observed plateaus.
Actually, non-diagonal solutions exist generically in RCFT and have been
much investigated in the past, revealing many connections with deep
aspects of mathematics \cite{ciz}.
There are two possible mechanisms which produce them \cite{mose}: 
i) the symmetry algebra of the RCFT
can be enlarged by using other chiral fields in the theory;
ii) there is an automorphism of the fusion algebra, which can be
used to pair non-trivially the left and right representations.

We do find non-diagonal partition functions in our problem,
which correspond to those found by Itzykson in the analysis of the $\suem$
theories \cite{itz}\footnote{Further solutions might exist.}.
For $m$ containing a square factor, $m=a^2m^\prime$, we find partition 
functions with an extension of the $\suem$ symmetry and a reduced set of 
excitations with respect to the diagonal Jain theories.
These new edge theories are in agreement with the experimental pattern
beyond the Jain series.
Moreover, there are modular invariants with a twist between charged and neutral
$\suem$ weights, which is due to an automorphism of the fusion rules. 
These invariants occur for filling fractions with large denominators,
which are not experimentally observable. 

The non-diagonal partition functions with extensions of the 
$\u1\otimes\widehat{SU\left(a^2 m^\prime \right)}$ symmetry 
span again the Jain filling fractions, with
$m\to m^\prime$, and further series of fractions which 
include $\nu=2/3,4/5,6/7,8/9,\dots$, $4/11, 4/13$ (only) 
for small values of the denominator.
We show a consistent interpretation of almost all the experimental
data \cite{survey}, which only contains these filling fractions and 
their (less stable) ``charge-conjugated'' partners $\nu\to (1-\nu)$. 
Therefore, our analysis of modular invariant partition functions 
is a valid alternative to the higher orders of the Jain hierarchy of 
wave functions \cite{jain} and similar constructions, which generically 
predict too many unobserved filling fractions beyond $\nu=4/11,4/13$.

In the conclusions, we remark that the modular invariant partition 
function and the associated RCFT calculus can be useful for understanding
the non-Abelian edge theories which have been discussed in 
Refs.\cite{mr}\cite{nonab}\cite{pfaff}\cite{esko}.
Actually, Ref.\cite{mr} already suggested the use of some 
RCFT properties which are explained in the present paper.
Moreover, partition functions for these theories have been recently
found in Ref.\cite{read}, whose general properties are in agreement with our 
analysis. 
Finally, in the Appendix A, we collect some useful formulas for the
theta functions occurring in the conformal characters.

%- 2 ------------------------------------------------------

\section{Partition functions for the Laughlin plateaus}
\label{sec2}

Let us consider the annulus geometry ${\rm I}\times S^1\times S^1$,
with coordinates $(r,\varphi, t_E)$,
for the plane and the Euclidean compact time, where
$r\in {\rm I}=[R_L,R_R]$ and $t_E=-it\equiv t_E+\beta$.
The electrons are supposed to form an incompressible fluid in the bulk
of the annulus, with gapful quasi-particle excitations.
We want to consider the partition function for the excitations
below the gap, which can be created in a conduction experiment
by attaching wires to the two edges.
These excitations are chiral and anti-chiral waves on the outer $(R)$
and inner $(L)$ edges, respectively. Their Hilbert space is made of pairs
of representations of the $\winf$ algebra of quantum area-preserving
diffeomorphisms of the incompressible fluid\footnote{
See Refs.\cite{cdtz1}\cite{ctz4}\cite{ctz5} for an introduction to the $\winf$ 
symmetry in the quantum Hall effect.}.
In this section, we consider the theories made by the simplest $\winf$
\reps , which are equivalent to those of the $\u1$ affine algebra
and have Virasoro central charge $c$ equal to one \cite{ctz5}.

Previous investigations \cite{juerg}\cite{cdtz1}
analysed the excitations of a single edge,
which are described by {\it chiral} CFTs. For these theories, the 
Virasoro eigenvalue $L_0$ gives the fractional spin $J$ which is also one-half
of the fractional statistics $\theta/\pi$ of the excitations of 
the incompressible fluid \cite{wilc}.
The following spectra were obtained for $J$, the filling fraction 
$\nu$ and the charge $Q$ :
\beq
\nu={1\over p}\ , \quad Q={n\over p}\ , \quad J=L_0={n^2\over 2p}\ ,
\qquad n \in \Z\ , \ p=1,3,5,\dots\ .
\label{specu1}\eeq
Therefore, these edge theories describe the quantum Hall effect at the
Laughlin plateaus. Each value of $(Q,L_0)$ is the weight of
a highest-weight \rep of the $\u1$ algebra, which contains a tower of
neutral excitations with quantum numbers
$(Q, L_0+k)$, $k>0$ integer (the descendant states) \cite{gins}.

Here, we shall derive independently the chiral-antichiral spectrum
on the annulus geometry, and find agreement with (\ref{specu1}) for
each of the two edges. We shall use the conformal field theory data,
impose the modular invariance of the partition function and
some other physical conditions.
The partition function \cite{cardy}\cite{ciz} is defined by
\beq
{\rm Z}\left(\tau,\zeta \right)\ = \ {\cal K}
\ {\rm Tr}\ \left[ {\rm e}^{i2\pi \left(
\tau\left(L_0^L- c/24\right) - \bar\tau\left(L_0^R- c/24\right) +
\zeta Q^L+ \bar\zeta Q^R \right) } \right]\ ,
\label{zdef}\eeq
where the trace is over the states of the Hilbert space, ${\cal K}$
is a normalisation to be described later and $(\tau,\zeta)$ are complex 
numbers. We recognise (\ref{zdef}) as a grand-canonical partition function,
with a heat bath and a particle reservoir: the operators
\beq
H_{CFT}={1\over R}\left(L^R_0+L^L_0- {c\over 12}\right)
+V_o\left(Q^L-Q^R\right) +\ {\rm const.},
\qquad J=L^L_0-L^R_0\ ,
\label{hconf}\eeq
measure the energy and spin of the excitations, respectively.
The real and imaginary parts of $(\tau, \zeta)$ are related to
the inverse temperature $\beta=1/k_BT= 2\pi R\ \I\tau >0$,
the ``torsion'' $\R\tau$, the chemical potential
$\mu/k_BT=2\pi \R\zeta$ and the electric potential between the
edges $V_o/k_BT=2\pi \I\zeta$.
Given that the Virasoro dimension is roughly the square of the charge,
the partition sum is convergent for $\I\tau >0$ and
$\zeta\in {\bf C}$.
We have naturally chosen a symmetric Hamiltonian for the two edges,
by adjusting the velocities of propagation of excitations,
$(v_L/R_L=v_R/R_R\equiv 1/R)$. The energy spectrum given by
$H_{CFT}$ is not completely realistic, because it does not include
the logarithmic correction due to the bulk Coulomb interaction among the 
electrons. However, we know that this correction would 
not change the form of the Hilbert space \cite{ctz4};
one can use the simpler $H_{CFT}$ to construct a ``kinematical''
partition function which describes the universality class
of the low-energy excitations, namely that is an inventory of their
quantum numbers and abundances.

As usual, we can divide the trace in (\ref{zdef}) into a sum over
pairs of highest-weight $\u1$ \reps and the sums over the states within
each \rep. The latter give rise to the $\u1$ characters \cite{gins}
\beq
Ch\left(Q,L_0\right)\ = \left.{\rm Tr}\right\vert_{\u1}\
\left[ {\rm e}^{i2\pi \left(
\tau\left(L_0- c/24\right) + \zeta Q\right) } \right]\ = \
{q^{L_0}\ w^Q \over \eta(q)}\ ,
\label{charu1}\eeq
where $\eta$ is the Dedekind function,
\beq
\eta\left( q\right) =q^{1/24}\ \prod_{k=1}^\infty \left( 1-q^k \right)\ ,
\qquad q={\rm e}^{\disp i2\pi \tau}\ ,\quad w={\rm e}^{\disp i2\pi \zeta}\ .
\label{dede0}\eeq
Any conformal field theory with $c\ge 1$ contains an infinity of
Virasoro (and $\u1$) representations \cite{cardy}; therefore, we must further
regroup the $\u1$ characters into characters $\chi$ of an extended
algebra in order to get the finite-dimensional expansion
\beq
Z\ =\ \sum_{\lambda,\bar\lambda=1}^N\
{\cal N}_{\lambda,\bar\lambda}\ { \chi}_{\lambda}\
{ \bar\chi}_{\bar\lambda}\ ,
\label{rcftz}\eeq
which characterises a rational conformal field theory \cite{mose}.
The coefficients ${\cal N}_{\lambda,\bar\lambda}$ are unknown
positive integers known as the {\it multiplicities} of the
excitations.
The matrix ${\cal N}$, and the parameters specifying the $(Q,L_0)$
spectrum are the unknown quantities to be determined by the following
conditions.

\subsection{Modular invariance conditions}
\label{mic}

The partition function describes the pairing of excitations
on the two edges to form physical excitations of the
entire sample, which can be measured in a conduction
experiment. These can only be electron-like and should have 
integer or half-integer spin:
\beq
T^2:\qquad Z\left(\tau +2,\zeta\right) \equiv
{\rm Tr}\ \left[\cdots {\rm e}^{\ i2\pi\ 2\left(L^L_0 -L^R_0\right)}
\right] = Z\left(\tau, \zeta\right)\ .
\label{tcond}\eeq

In (\ref{zdef}), $Z(\tau)$ is the partition function
of a $(1+1)$-dimensional conformal field theory, where
$\tau$ is the modular parameter of a space-time torus.
On the other hand, (\ref{zdef}) is also the path-integral amplitude of the
Abelian Chern-Simons theory on the manifold
${\cal M}=S^1\times S^1\times {\rm I}$.
Actually, by exchanging the Euclidean time with one space coordinates, 
such that $t_E\in {\rm I}$, this is also an amplitude for compact space, 
which has been considered by Witten and other authors \cite{jones}\cite{c-s}.
They have shown that the Hilbert space of the Chern-Simons theory
on a space torus is described by a pair of $\u1$ affine algebras with
opposite chiralities, which is precisely our setting.
In both space-space and Euclidean space-time tori,
we can exchange the two periods
by a corresponding renaming of the parameters \cite{cardy}:
\beq
S:\qquad Z\left(-{1\over\tau},-{\zeta\over\tau}\right)=
Z\left(\tau,\zeta\right)\ .
\label{scond}\eeq

The transformations $ST^2S$ and $S$ generate the subgroup $\Gamma (2)$
of the modular group $\Gamma=SL(2,\Z)/\Z_2$ of rational
transformations $\tau \to (a\tau +b)/(c\tau +d)$, $a,b,c,d\in \Z$,
which are subjected to the conditions $(a,d)$ odd and $(b,c)$ even.
Thus, $Z$ is invariant under the modular subgroup $\Gamma (2)$,
due to the presence of one-fermion states, as it 
occurs in the Neveu-Schwarz sector of the supersymmetric 
theories\cite{gins}\footnote{
Actually, $S$ and $T^2$ generate a slightly larger subgroup of $\Gamma$;
see appendix A.}.

We now discuss two further conditions concerning the charge spectrum.
In general, the incompressible fluid contains anyons,
which are vortices in the bulk of the fluid, spilling a fractional charge
to the edge. They have a finite gap $O(e^2/\ell)=O(e^2\sqrt{eB})\gg eV_o$, 
due to the Coulomb interaction \cite{laugh}, and cannot be generated in a 
conduction experiment. They can be produced by tuning the magnetic field 
away from the center of the plateaus, becoming subsequently pinned down by 
impurities and turning into static, fractional charges \cite{prange}. 
The edge excitations in presence of them
are described by the expectation value of Wilson lines in the Chern-Simons 
theory \cite{jones}, while the on-centered situation without bulk anyons
is described by the partition function (\ref{zdef}). In the latter case,
the edge excitations should have total integer charge,
$Q^L+Q^R \in \Z$, which is the number of electrons carried to the edges
by the wires. This condition is enforced by:
\beq
U:\qquad Z\left(\tau,\zeta +1\right) \equiv
{\rm Tr}\ \left[\cdots {\rm e}^{\ i2\pi\left(Q^R +Q^L\right)}
\right] = Z\left(\tau, \zeta\right)\ .
\label{ucond}\eeq

Owing to this condition, there can be fractionally charged excitations
at one edge, but these should pair with complementary ones on the other
boundary. Consider, for example, $\nu=1/3$; we can imagine to drop in an
electron, which splits into the pair $(Q^L,Q^R)=(1/3,2/3)$, or another one
among $(0,1),(2/3,1/3),(1,0)$. These different splittings should all be 
realizable by appropriate tuning of the potential $V_o$ in (\ref{hconf}).
Their equivalence gives a symmetry of the partition function;
the shift among them is called {\it spectral flow} \cite{cdtz1}\cite{ctz5}.
This flow has been discussed by Laughlin in the thought experiment defining 
the fractional charge on the annulus \cite{laugh}: the addition of a quantum
of flux through the center of the annulus is a symmetry of the
gauge-invariant Hamiltonian but causes a flow of all the quantum states
among themselves. We can simulate this flow by requiring invariance of 
the partition functions under a shift of the electric potential 
by a corresponding
amount (in our notations $e=c=\hbar=1$) $V_o \to V_o+1/R$, namely
$\zeta\to\zeta +\tau$:
\beq
V:\qquad Z\left(\tau,\zeta +\tau\right) = Z\left(\tau, \zeta\right)\ .
\label{vcond}\eeq
Note that the transport of an elementary fractional charge between the two
edges is related to the Hall conductivity, thus the spectral
flow also determines the Hall current.

The conditions $(T^2, U, V)$ have been motivated by specific properties
of the quantum Hall effect. However, analogous conditions
are found in the canonical quantization of the Abelian
Chern-Simons theory on the space torus \cite{c-s}: the Chern-Simons field
$A_i, i=1,2$ is a flat connection,
which has two non-trivial degrees of freedom on the torus,
$(\theta_1(t),\theta_2(t))$. One identifies our variable
$\zeta\sim \theta_1+\tau\theta_2$; furthermore, the conditions $(U,V)$ 
correspond to the Gauss law which enforces gauge invariance on physical states.

We now solve the conditions $(T^2,S,U,V)$ for the $c=1$ theory.
Owing to the condition $U$, we can first collect left $\u1$ \reps
which have integer spaced charges $Q^L=\lambda/p+\Z$,
$p$ integer, and then combine them with the corresponding right \reps .
The Virasoro dimensions are given by Eq.(\ref{specu1}), without loss
of generality; the charge unit $1/p$ is free and related
to the normalisation of $\zeta$.
The sums of $\u1$ characters having integer spaced charge should give 
the characters of the extended algebra $\chi_\lambda$ (\ref{rcftz}),
which carry a finite-dimensional representation of the
modular group \cite{mose}. These are known to be 
the theta functions with rational characteristics\footnote{
See the Appendix A for more details on modular functions and transformations.} 
$\Theta\left[{a\atop b}\right]\ $:
\barr
\chi_\lambda &=&
{\rm e}^{\disp -{\pi\over p}{\left(\I\zeta\right)^2\over \I\tau} }
\ {1\over \eta}\ \Theta\left[{ \lambda/p \atop 0}\right]
\left(\zeta\vert p\tau\right) \nonumber\\
&=& {\rm e}^{\disp -{\pi\over p}{\left(\I\zeta\right)^2\over \I\tau} }
\ {1\over \eta}\ \sum_{k\in \Z}\
{\rm e}^{\disp\ i2\pi \left(\tau{(pk+\lambda)^2\over 2p} +
\zeta\left({\lambda\over p}+k\right)\right) }\ ,
\label{thetaf}\earr
which carry charge $Q^L=\lambda/p + {\bf Z}$, $\lambda=1,2,\dots,p$. 
The prefactor will be explained later.
The transformations $T^2,S,U,V$ of these generalised characters are:
\barr
T^2:\ \chi_\lambda\left(\tau+2,\zeta\right)\ & = &
{\rm e}^{\disp\ i2\pi\left({\lambda^2\over p}-{1\over 12} \right) }
\chi_\lambda\left(\tau,\zeta\right)\ ,\\
S:\ \chi_\lambda\left(-{1\over\tau},-{\zeta\over\tau}\right) & = &
{ {\rm e}^{\disp\ i{\pi\over p}\R {\zeta^2 \over \tau} } \over\sqrt{p}}
\ \sum_{\lambda^\prime=0}^{p-1}\
{\rm e}^{\disp i2\pi{\lambda\lambda^\prime\over p} } \
\chi_{\lambda^\prime}\left(\tau,\zeta\right)\ ,\\
U:\ \chi_\lambda\left(\tau,\zeta+1\right)\ & = &
{\rm e}^{\disp\ i2\pi\lambda/ p } \
\chi_\lambda\left(\tau,\zeta\right)\ ,\\
V:\ \chi_\lambda\left(\tau,\zeta+\tau\right)\ & = &
{\rm e}^{\disp -i{2\pi\over p}\left(\R\zeta +\R{\tau\over 2} \right) }
\chi_{\lambda+1}\left(\tau,\zeta\right)\ .
\label{chitr}\earr
These transformations show that the characters $\chi_\lambda$ carry a
unitary \rep of the modular group $\Gamma(2)$, which is projective
for $\zeta\neq 0$ (the composition law is verified up to a phase).

The corresponding sums of $\u1$ \reps for the other edge are given by
$\bar\chi_{\bar\lambda}$ carrying charge $Q^R=-\bar\lambda/p+{\bf Z}$.
The $U$ condition (\ref{ucond}) is applied to $Z$ in the form 
(\ref{rcftz}) and it requires that left and right charges satisfy
$\lambda=\bar\lambda $ mod $p$. The unique\footnote{
This will be proven in section 4.3.} solution
is $\lambda=\bar\lambda$, which also satisfies the $(T^2,S,V)$ 
conditions, leading to the invariant:
\beq
Z=\sum_{\lambda=1}^p\ \chi_\lambda\ \bar\chi_{\lambda} \ .
\label{zedone}\eeq
%}
This is the annulus partition function for the $c=1$ edge
theories we were after. It yields the multiplicities of \reps 
${\cal N}_{\lambda,\bar\lambda}= \delta^{(p)}_{\lambda,\bar\lambda}$,
and still depends on two free parameters,
$p$ and $\zeta$, which are determined by further physical conditions.
These have been discussed in the literature \cite{zee}\cite{ctz3}\cite{ctz5} 
and are simply reformulated in the present context:
\begin{itemize}
\item
The normalisation of $\zeta$, i.e. of the charge
unit, is determined self-consistently as follows.
Among the transformations (\ref{chitr}),
$V$ is the only one sensitive to rescalings $\zeta\to a\zeta$,
with $a$ integer: we fix $a=1$ by requiring that the minimal
spectral flow $\zeta\to\zeta+\tau$ carry the minimal amount of
fractional charge $(1/p)$ from one edge to the other,
$\chi_\lambda(\zeta+\tau)\to\chi_{\lambda+1}(\zeta)$.
This corresponds to the definition of the (fractional) charge
in the Laughlin thought experiment \cite{laugh}.
Moreover, the amount of displaced charge per unit of flux
is a measure of the Hall conductivity: we conclude that
(\ref{zedone}) describes the Hall effect at filling fractions $\nu=1/p$.
\item
Within the spectrum of charges and fractional spins of
$Z$ (\ref{zedone}), there should be electron states on each edge
which have unit charge and odd statistics, $2J=1$ mod $2$.
{}From (\ref{specu1}) one can see that this requirement fixes $p$
to be an odd integer.
Therefore, these partition functions describe the Laughlin plateaus
$\nu=1,1/3,1/5,\dots$
\item
One can also verify that all the excitations have
integer monodromies with respect to the electrons:
\beq
J[n_e]+J[n]-J[n_e+n] \in \Z\ ,
\label{estat}\eeq
where $(n_e=p,n)$ denote the integer labels in (\ref{specu1}) for the 
electron and a generic excitation, respectively.
\end{itemize}

Let us add more comments about the main result of this section,
Eqs.(\ref{zedone}),(\ref{thetaf}).
The non-holomorphic prefactor added to the characters in (\ref{thetaf})
is the constant term in the Hamiltonian
(\ref{hconf}), appropriately tuned to have the spectrum
\beq
E_{n_L,n_R} ={1\over R}{1\over 2p}\left[
\left(n_L+RV_o\right)^2\ +\ \left(n_R-RV_o\right)^2 \right]\ ,
\label{tunspec}\eeq
whose minimum is independent of the the value of $V_o$.
Actually, this is necessary for the invariance of $Z$ under the
spectral flow, and it amounts to adding a capacitive energy
to the edges equal to $E_c=RV_o^2/2p$.
Note that this prefactor also appears in the quantization of
the Chern-Simons theory on the space torus, in the measure
for the wave-functions inner product \cite{c-s}.

A more explicit form of the partition function (\ref{zedone}) is
\beq
Z={{\rm e}^{\disp -{2\pi(\I\zeta)^2\over p\ \I\tau} }
            \over\left\vert\eta\right\vert^2 }\
\sum_{\lambda=0}^{p-1}\ \sum_{k,\bar k\in \Z}\
q^{\disp {\left(pk+\lambda\right)^2\over 2p} }\
\bar q^{\disp {\left(p\bar k+\lambda\right)^2\over 2p } }\
w^{\disp {pk+\lambda\over p} }\
\bar w^{\disp {p\bar k+\lambda\over p}}\ .
\label{zexp}\eeq
This expression is {\it not} the same as the well-known
partition function of a real bosonic field compactified
on a rational circle of radius $R_c=r/2s$ \cite{gins},
\beq
Z_B={1\over\left\vert\eta\right\vert^2}\
\sum_{n,m\in \Z}\
q^{\disp {1\over 2}\left( {n\over 2R_c} +mR_c\right)^2 }\
\bar q^{\disp {1\over 2}\left( {n\over 2R_c} -mR_c\right)^2 }\ .
\label{zbose}\eeq
Indeed, after setting $\zeta=0$, $Z$ and $Z_B$ would be
equal for $R_c=1/p$, which would require $p$ {\it even}.
Therefore, the RCFT of the Laughlin plateaus is an {\it odd}
modification of the rational compactified boson, which is
invariant under the subgroup $\Gamma (2)$ of the modular group.

\subsection{Modular invariance and the fusion algebra}
\label{fusal}

We have shown that the partition function for the Laughlin
fluids on the annulus takes the standard form of rational
conformal field theories (RCFT) \cite{mose} and is modular invariant
($S$ invariant, in particular).
This implies that we can carry over to the quantum Hall effect
some interesting properties of the RCFT \cite{mr}, due to Verlinde 
\cite{verl}, Witten \cite{jones} and others \cite{mose},
relating the fusion rules, the modular transformation $S$ and the dimension 
of Hilbert spaces of the Chern-Simons theories. 
In particular, we shall use the fusion rules to compute the
Wen ``topological order'' \cite{topord} of the quantum Hall effect in any
geometry and in presence of impurities.

Let us first recall the Verlinde theory relating the fusion rules to
the $S$ transformation \cite{verl}\cite{mose}.
The generalised characters $\chi_\lambda$ of the RCFT
(\ref{thetaf}) correspond to the  extension of the $\u1$ algebra by a
chiral current $J_p$. This is a $\u1$ \rep which enters
in the vacuum \rep of the extended algebra, and then it
can be identified from the first non-trivial term in the expansion of
$\chi_0$ into $\u1$ characters: $J_p$ has half-integer dimension $L_0=p/2$
and unit charge\footnote{Similar to the supersymmetric partner of the
stress tensor $T_F$ ($L_0=3/2$) \cite{gins}.}.
The highest-weight \reps of the RCFT algebra correspond
to generalised primary fields $\phi_\lambda$, whose charge
is defined modulo one (the charge of $J_p$); namely, they
collect all chiral excitations with the same fractional charge.

The fusion rules \cite{gins} are the selection rules for the
operator-product expansion (in simple terms, the making
of composite excitations), and are an associative Abelian algebra of the
fields. The $\u1$ fusion rules are simply given by charge conservation;
thus, those of the extended algebra are given by
the additive group $\Z_p$ of integers modulo $p$,
\beq
\phi_i \cdot \phi_j \equiv N_{ij}^k \ \phi_k\ , \qquad
N_{ij}^k = \delta^{(p)}_{i+j,k}\ , \quad i,j,k \in \Z_p\ .
\label{erik}\eeq
One can use the conjugacy matrix  $C_{ij}=N_{ij}^0$ as a metric
to raise and lower indices: then $N_{ijk}=\delta^{(p)}_{i+j+k,0}$
is a completely symmetric tensor.
The $N\times N$ matrices $\left(N_i\right)_j^k=N_{ij}^k$ give a
representation of the fusion algebra
which decomposes into $N$ one-dimensional irreducible ones,
represented by the eigenvalues $\lambda_i^{(n)}$,
$n=0,\dots,p-1$ of the matrix $N_i$.
Verlinde has shown that these eigenvalues are related to
the matrix elements of the modular transformation $S_k^n$, as follows:
\beq
\lambda_i^{(n)}={S_i^n\over S_0^n}\ ,\qquad {\rm and}\quad
N_i=S\Lambda_iS^\dagger, \quad
\left(\Lambda_i\right)_n^k=\delta_n^k\ \lambda_i^{(n)} \ .
\label{opes}\eeq
One immediately verifies these formulae for the case of the annulus partition
function (\ref{zedone}) by using the representation (\ref{chitr})
for $\zeta=0$ : $S_k^n=\exp(i2\pi\ kn/p)/\sqrt{p}$.
Hereafter we consider the specialised partition functions 
$Z\left(\tau,\zeta=0\right)$, which have been analysed by Verlinde. 

A general property of observables in conformal field theory is that they
decompose into a finite sum of chiral-antichiral terms,
called conformal blocks, which are determined by 
representation-theoretic properties \cite{gins}. 
For example, $Z$ decomposes into pairs of characters.
The space of conformal blocks ${\cal H}$ for the $n$-point functions on a
genus $g$ Riemann surface $\Sigma_g$ is a vector bundle, whose
vectors are labelled by the  weights of the \reps of the symmetry algebra
\cite{mose}\cite{verl}.
A nice application of the fusion rules is to compute the dimension of
this space. Clearly, a \rep index $i_k$ is associated to each of the
$n$ points, or ``punctures'' on $\Sigma_g$.
The coefficients $N_{ij}^k$ represent the three-point
function on the sphere $\Sigma_0$; we can associate to it a ``dual''
$\varphi^3$ vertex made by three oriented lines joined to a point.
Higher genus surfaces can be built by sewing three-punctured spheres:
for example, a handle is made by joining two punctures
and summing over \reps with the ``propagator'' $\delta_i^k$ or
$C_{ik}$, according to the indices of the punctures,
i.e. to the line orientations in the $\varphi^3$ graph.
There are many ways to associate a $\varphi^3$ graph to a Riemann
surface, which are related by the duality symmetry, a fundamental
property of the fusion rules (\ref{erik}).
For example, let us compute the number of
blocks occurring in the one-point functions on the torus
$\langle\phi_i\rangle_T$ for the Abelian ${\bf Z}_p$ fusion rules:
\beq
dim{\cal H}\left(\Sigma_1;(P,i)\right)=\sum_{k=1}^p\ N_{ik}^k=
\sum_{k=1}^p\delta^{(p)}_{i,0}=\ p\ \delta^{(p)}_{i,0}\ .
\label{1pi}\eeq
Namely, all torus one-point functions vanish (by charge conservation)
and the torus partition function $\langle{\bf 1}\rangle_T$
decomposes in $p$ blocks. Note that the same number of terms was found
for $Z$ (\ref{zedone}) computed before.

The general formula for the dimension of the space of conformal blocks on 
a genus $g$ surface with $n$ punctures was found in Ref.\cite{mose}; 
owing to the relations (\ref{opes}), this can be expressed in terms of the 
$S$ matrix elements, as follows:
\beq
dim{\cal H}\left(\Sigma_g;(P_1,i_1),\dots,(P_n,i_n)\right)=
\sum_k\ \left({1\over S_0^k}\right)^{2(g-1)}\
{S_{i_1}^k\over S_0^k}\cdots{S_{i_n}^k\over S_0^k}\ .
\label{dimhg}\eeq

We now recall the two Witten correspondences \cite{jones}\cite{c-s} between 
the RCFT based on the affine Lie algebra $\widehat G$ (the Wess-Zumino-Witten
model) and the Chern-Simons theory with gauge group $G$ canonically 
quantized on a manifold ${\cal M}=\Sigma\times {\bf R}$, where
{\bf R} is the time axis:
\begin{itemize}
\item
If $\Sigma$ has a boundary, for example a disk,  
the quantization of the Chern-Simons gauge field gives rise to 
the Wess-Zumino-Witten model on $\partial {\cal M}$.
This is the familiar relation for the CFT description of 
edge excitations in the quantum Hall effect \cite{wen}.
On any boundary circle, we have the spectrum of edge excitations described
by an affine highest-weight \rep; its weight is free, but 
the global conservations laws on the surface give a constraint.
\item
If $\Sigma_g$ is compact, the Chern-Simons gauge field is a 
non-trivial flat connection which has a finite
number of degrees of freedom. The finite-dimensional Hilbert space is 
equivalent to the space of conformal blocks ${\cal H}(\Sigma_g)$ 
of the Euclidean RCFT defined on $\Sigma_g$:
the $n$-point conformal block of the RCFT corresponds to the
Chern-Simons wave functional for $n$ time-like Wilson lines piercing 
the surface.
\end{itemize}

Form the second relation, we conclude that the formula (\ref{dimhg}) 
gives the dimension of the Chern-Simons Hilbert space on $\Sigma_g$
with $n$ Wilson lines.
Furthermore, the surfaces $\Sigma_{g,k}$ 
with $k$ boundaries are relevant for the quantum Hall effect, 
and we need to extend the formula (\ref{dimhg}) to them.
For what concerns the counting of states, a disk
is equivalent to a puncture whose quantum numbers are not fixed a priori.
This fits with the simple-minded observation that, topologically,
one can introduce a boundary on a closed surface by removing a disk from it.
Therefore, the dimension formula for ${\cal H}(\Sigma_{g,k})$
with boundaries is obtained from those for ${\cal H}(\Sigma_g)$ in 
(\ref{dimhg}), by adding $k$ punctures, and by summing over the 
corresponding free indices $\{i_1,\dots,i_k\}$.

\subsection{The topological order and its independence of impurities}
\label{stopo}

Wen has introduced and stressed the concept of topological order in
the quantum Hall effect \cite{topord}; by this he means a number of
properties which can be summarised by saying that the incompressible
fluid ground state is not a single quantum state but a topological
field theory, which is actually the Chern-Simons theory.
Here, the topological order will specifically mean 
the {\it number} of degenerate ground states on a closed surface 
$\Sigma_g$. This degeneracy is hard to derive from the microscopic 
dynamics of the electrons, but it is indeed present in the trial 
wave functions and in numerical spectra on a toroidal 
space \cite{hr}.

{}From the previous discussion, we understand that these
degenerate ground states are described by the Hilbert space 
${\cal H}(\Sigma_g)$ of the Chern-Simons theory.
Therefore, we can compute the topological order by using the previous
counting formula (\ref{dimhg}). For the torus, we find: 
\beq
dim{\cal H}\left(\Sigma_1\right)= \sum_{k=1}^N\ 1\ = N\ ,
\qquad {\rm for \ any\ RCFT}\ ,
\label{topto}\eeq
where $N$ is the dimension of the \rep of the $S$ transformation, i.e. the 
number of \reps of the (maximally) extended symmetry algebra of the RCFT.

The same topological order can be computed from the annulus partition
function introduced before.
Actually, consider the dimension formula for the Chern-Simons
theory on ${\rm I}\times S^1\times S^1$. The compact time evolution
is not important, so this is equivalent to
${\rm I}\times S^1\times {\bf R}$. The annulus surface is topologically
equivalent to the sphere with two disks removed, i.e. with two punctures.
We use again the dimension formula (\ref{dimhg}) adding the boundaries,
and we find:
\barr
dim{\cal H}\left(\Sigma_{0,2}\right) &=&
\sum_{i,j=1}^N \ dim {\cal H}\left(\Sigma_0,(P,i),(P^\prime, j)\right)
\nonumber\\
&=& \sum_{i,j}\ N_{ij,0}=\sum_{i,j}\ C_{i,j}= \ N\ ,\qquad
{\rm for\ any\ RCFT}\ .
\label{topan}\earr
{}From this and previous results, we conclude that 
the annulus partition function in section $(2.1)$ gives a correct
description of the standard RCFT properties and it encodes the
Wen topological order, previously defined for the quantum Hall effect on
compact spaces \cite{topord}.

In the literature of the quantum Hall effect, both Witten relations
have been considered, resulting into a ``bulk'' description of the 
incompressible Hall fluids by an Euclidean RCFT \cite{bulk}
and an ``edge'' RCFT description, as in the present paper.
The two descriptions and conformal field theories are equivalent \cite{mr}: 
the second one has a more immediate physical meaning and is often
simpler. It is rather useful that the important concept of topological
order is shown by observables of the edge conformal field theory.

The topological order on higher genus surfaces can be similarly computed
from (\ref{dimhg}): one needs the specific form of the $S$ transformation
and the result is characteristic of each theory.
For the Abelian Laughlin fluids, we use the ${\bf Z}_p$  $S$-matrix
(\ref{chitr}) and obtain
\beq
dim{\cal H}\left(\Sigma_g\right)=
\sum_{k=1}^p \ p^{\ g-1} = \ p^{\ g}\ , \qquad \nu={1\over p}\ ,
\ p=1,3,\dots
\label{topo}\eeq
This agrees with the explicit calculation in the Abelian Chern-Simons
theory \cite{topord}.

Another interesting application of the Verlinde calculus is to
compute the topological order in presence of impurities.
Let us again consider the quantum Hall effect on the annulus, and
assume there exists $n$ impurities in the bulk, i.e. $n$ anyons.
These are static and do not contribute to the Hall conduction \cite{prange}; 
as discussed before, they modify the ground state for the edge excitations,
which are no longer described by the partition function (\ref{rcftz}).
For example, the charge balance between the edges (\ref{ucond}) is
modified to $Q^L+Q^R+Q_{\rm bulk}\in {\bf Z}$.
These edge excitations are described by the Chern-Simons expectation value
of $n$ time-like Wilson lines piercing the annulus \cite{c-s}.
Contrary to the partition function, this expectation value is 
not modular invariant because the Wilson lines select a time direction.
Nevertheless,we can easily find the corresponding topological order
by using again (\ref{dimhg}). The $n$ punctures on the
annulus correspond to $n+2$ punctures on the sphere, with sum
over the indices of the two additional ones; for the
Laughlin plateaus, we use again the ${\bf Z}_p$ $S$ transformation and find:
\beq
dim{\cal H}\left(\Sigma_{0,2};(P_1,i_1),\dots,(P_n,i_n)\right) =
\sum_{i,j=1}^p\ \delta^{(p)}_{i+j+i_1+\cdots +i_n,0}= \ p \ ,
\quad \nu={1\over p}\ .
\label{topinp}\eeq
This shows that the topological order on the annulus is
independent of impurities, for Abelian fluids.
Similarly, one verifies that the topological order 
on $\Sigma_g$ is also independent of impurities\footnote{
The added anyons must have total vanishing charge on a compact surface.}
in agreement with the perturbative analysis of Ref.\cite{topord}.
The independence of impurities is a rather important result;
it shows that the topological order is robust, since it holds
in generic experimental situations.
On the other hand, the partition function is modular invariant only
in absence of impurities, although it encodes many universal properties
which are independent of them.

%- 3 ------------------------

\setcounter{footnote}{1}

\section{$\winf$ minimal and non-minimal theories}
\label{sec3}

In this section, we recall some properties of the conformal
field theories which have been proposed to describe the Jain series of 
stable plateaus at $\nu=m/(ms\pm 1)$, $m=2,3,\dots$ and $s>0$ even integer.
These are of two types: 
\begin{itemize}
\item The minimal models of the $\winf$ algebra \cite{ctz5};
\item Their non-minimal version, the theories with 
$\u1\otimes\suem$ affine algebra \cite{juerg}\cite{zee}.
\end{itemize}
Both types of theories possess the $\winf$ dynamical symmetry of
area-preserving diffeomorphisms, which expresses at the quantum level
the semiclassical property of incompressibility of the electron fluid 
at the plateaus \cite{ctz1}\cite{sakita}\cite{ctz3}. 
These theories have been constructed in detail in Ref.\cite{ctz5}
by using the representation theory of the $\winf$ algebra developed in 
Refs.\cite{kac1}\cite{kac2}.

Historically, the non-minimal theories were first introduced as a
multi-component generalisation of the Abelian theory of
the previous section \cite{juerg}. 
Actually, the $m$-dimensional generalisation
of the Abelian spectrum (\ref{specu1}) is:
\barr
Q & = & \sum_{i,j=1}^m\ K^{-1}_{ij}\ n_j\ , \qquad n_i\in {\bf Z} \nl
J & = & {1\over 2}\ \sum_{i,j=1}^m\ n_i\ K^{-1}_{ij}\ n_j\ , \nl
\nu & = & \sum_{i,j=1}^m\ K^{-1}_{ij}\ , \qquad c=m\ ,
\label{mabel}\earr
where $K_{ij}$ is an arbitrary symmetric matrix of couplings,
with integer elements, odd on the diagonal.
The derivation of (\ref{mabel}) goes as follows: one assumes that the 
electron fluid has
$m$ independent layers - naively, one could make an analogy with the 
integer Hall effect, with $m$ filled Landau levels, each one 
having an edge described by an Abelian current, altogether
yielding the $\u1^{\otimes m}$ affine algebra \cite{gins}. 
Its highest-weight  \reps are labelled by a vector 
of (mathematical) charges $r_a, a=1,\dots,m$,
which spans an $m$-dimensional lattice ${\bf r}=\sum_i{\bf v}^{(i)} n_i$, 
$n_i \in {\bf Z}$ as required by the closure of the fusion rules, given by
the addition of charge vectors \cite{ctz5}.  
The physical charge is a linear functional of ${\bf r}$ and the Virasoro
dimension is a quadratic form, both parametrised by the 
metric\footnote{
Lorentzian signature corresponds to excitations with mixed chiralities 
\cite{ctz5}.}
of the lattice $K^{-1}_{ij}\sim {\bf v}^{(i)}\cdot{\bf v}^{(j)}$.
This matrix $K$ is further constrained to take integer values by the 
requirement that the spectrum contains $m$ electron-like excitations
\cite{zee}. Therefore, each $m$-plet of integers $n_i$ in (\ref{mabel})
gives a highest-weight state of the $\u1^{\otimes m}$ algebra,
which is physically interpreted as an anyon excitation in the 
multi-layered electron fluid.

The spectrum (\ref{mabel}) is very general, due to the many free
parameters in the $K$ matrix. Actually, these can be chosen to 
comply with the results of all the known hierarchical constructions of wave
functions. In particular, the Jain hierarchy \cite{jain} 
was shown to correspond \cite{zee}
to $K_{ij}= \pm\delta_{ij} + s\ C_{ij}$, where 
$s>0$ is an even integer and $C_{ij}=1,\ \forall\ i,j=1,\dots,m$.
The corresponding spectrum is
\vbox{
\barr
\nu& =& {m\over ms \pm 1}\ , \qquad s >0 \ {\rm even\ integer}\ ,
\qquad c = m\ , \nonumber\\
Q & = & {1\over ms \pm 1}\ \sum_{i=1}^m n_i \ ,\nonumber\\
J &=& \pm {1\over 2}\left( \sum_{i=1}^m n^2_i -
{s\over ms \pm 1}\left( \sum_{i=1}^m n_i \right)^2 \right) \ .
\label{wspec}\earr
}
This spectrum is rather peculiar because it
contains $m(m-1)$ {\it neutral} states with unit Virasoro dimension 
$(Q,L_0)=(0,1)$. By using an explicit free bosonic field
construction, one can show \cite{juerg} that these are chiral currents 
$J_{\vec\beta}$, where ${\vec\beta}$ is a root of $SU(m)$ \cite{wyb}, which 
generate the $\u1\otimes\suem$ affine algebra, together with
the $m$ original $\u1$ currents for the Cartan subalgebra \cite{gins}.
The $\suem$ affine algebra is a RCFT with $m$ highest weight
representations, each one resumming a sector of the $\u1^{\otimes m}$
spectrum (\ref{wspec}).

The minimal models were obtained from the analysis 
of the $\winf$ symmetry \cite{ctz5}. The Laughlin semiclassical picture of 
the incompressible fluid of electrons was assumed.
At constant electron density and number, all the configurations
of the incompressible fluid have the same area; thus, they
can be mapped into each other by area-preserving diffeomorphisms
of the plane, obeying the classical $w_\infty$ algebra \cite{ctz1}
\cite{sakita}. The infinitesimal deformations are the edge excitations.
Their quantization leads to a conformal field theory with quantum algebra
$\winf$, which contains $\u1$ and Virasoro as subalgebras \cite{ctz3}.
We refer to our previous works for the description 
of this semiclassical picture and the definition
of the $\winf$ algebra, and we recall here the main properties of 
these theories \cite{ctz5}.

The $\winf$ theories can be constructed by assembling highest-weight
\reps of the $\winf$ algebra which are closed under their fusion rules.
The unitary \reps have integer central charge $(c=m)$ and can be of 
two types: {\it generic} or {\it degenerate} \cite{kac1}\cite{kac2}.
The {\it generic} \reps are one-to-one equivalent to those
of the Abelian affine algebra $\u1^{\otimes m}$ with charge vectors $r_a$
satisfying $(r_a-r_b) \not\in {\bf Z} \ ,\forall\ a\neq b$.
Clearly, the $\winf$ theories made of generic representations 
correspond to the generic  hierarchical Abelian theories in (\ref{mabel}).
On the other hand, the (fully) degenerate $\winf$ \reps are not equivalent
to $\u1^{\otimes m}$ \reps , rather are contained into them.
They are labelled by a charge vector ${\bf r}$  
satisfying  $(r_a-r_b) \in {\bf Z} \ ,\forall\ a\neq b$: 
an Abelian \rep with the same charge vector decomposes
into many degenerate $\winf$ representations with vectors ${\bf r} +$
(integers).
One can construct edge theories which are made of degenerate \reps only, 
which are called $\winf$ {\it minimal models} \cite{ctz5}.
The main result is that these are in one-to-one
relation with the Jain series of filling fractions and yield the same
spectrum (\ref{wspec}) of charge and fractional spins 
of the Abelian theories, subjected to the further condition 
$n_1\ge n_2\ge\dots\ge n_m$.
Basically, the $\winf$ degeneracy conditions on the vector ${\bf r}$
select uniquely the lattices corresponding to the ``good''
$K$ matrices for the Jain spectrum.

The detailed predictions of the $\winf$ minimal models
are however different from those of the Abelian theories \cite{ctz5}:
\begin{itemize}
\item 
There is a {\it single}, as opposed to $m$, independent Abelian 
currents, and, therefore, a single (fractionally) charged elementary 
excitation; the neutral excitations cannot be associated to $(m-1)$
independent edges. 
\item 
The neutral excitations have a $SU(m)$ (not $\widehat{SU(m)}_1$) 
associated ``isospin'' quantum number, given by the highest weight,
\beq
{\bf \Lambda}=\sum_{i=1}^{m-1}\ {\bf \Lambda}^{(i)}\ \left(
n_i - n_{i+1} \right) \ ,
\label{sumwei}\eeq
where ${\bf \Lambda}^{(i)}$ are the fundamental weights of $SU(m)$ 
\cite{wyb} and $n_i$ are the integers in Eq.(\ref{wspec}).
Actually, the degenerate $\winf$ \reps are equivalent to \reps
of the $\u1\otimes{\cal W}_m$ algebra, where ${\cal W}_m$ is the
Zamolodchikov-Fateev-Lykyanov algebra at $c=m-1$ \cite{fz}. 
The fusion rules of this algebra are isomorphic to
the tensor product of \reps of the $SU(m)$ Lie algebra; thus
the neutral excitations in these theories are quark-like and their
statistics is non-Abelian.
For example, the edge excitation corresponding to the electron is a composite
state, carrying both electric charge and the isospin  of the fundamental
\rep $\{ {\bf m}\}$ of $SU(m)$.
\item 
The number of particle-hole excitations above the ground state
is smaller in the $\winf$ minimal models than in the Abelian ones,
due to the corresponding inclusion of highest-weight \reps. 
\end{itemize}
These results show that the edge theories of the Jain plateaus 
possess rather non-trivial algebraic properties. The algebra inclusions are
\beq
\u1\otimes\suem\ \supset\ \u1^{\otimes m}\ \supset\ \u1\otimes{\cal W}_m \ .
\label{inclu}\eeq
The $\u1\otimes\suem$ theories are non-minimal 
realizations of the $\winf$ symmetry, which are possible precisely
at the Jain filling fractions.
More detailed properties can be better understood by an
explicit discussion of the simplest case $m=2$.

\subsection{The $c=2$ case}

The $\winf$ minimal models are made of representations of the
$\widehat{U(1)}\otimes {\rm Vir}$ algebra, because
${\cal W}_2$ is simply the $c=1$ Virasoro algebra.
The reduction of states from
the non-minimal to the minimal theories can be derived from
the characters \cite{gins} of the corresponding \reps \cite{ctz5}.
Neglecting the common $\u1$ factor, we compare the
characters of the $\su2$, $\u1$ and Virasoro \reps.
The charge vector of the $c=2$ degenerate $\winf$ \rep is
${\bf r}=\{r+n,r\}$, with $r\in {\bf R}$ and $n$ positive integer; 
the $c=1$ Virasoro part has the weight $L_0=n^2/4$ and the character
\beq
\chi^{\rm Vir}_{n^2/4} = q^{n^2/4} \left(1-q^{n+1}\right)/
\eta(q)\ .
\label{chiva}\eeq
These Virasoro \reps have associated an $SU(2)$ isospin quantum
number $s=n/2>0$, because their fusion rules are equivalent
to the addition of (total) spins 
$\{s+s^\prime\}\oplus\{s+s^\prime-1\}\oplus\dots\oplus\{|s-s^\prime|\}$.
The character of the $\u1$ algebra with the same weight $L_0$ is
\beq
\chi^{\u1}_{n^2/4} = q^{n^2/4} / \eta(q)\ .
\label{chiu}\eeq
Note that the Virasoro and $\u1$ characters differ by a
negative term in the numerator, which cancels part  of
the power expansion of $\eta(q)$.
Thus, the number of states above the highest-weight state is lower
in the Virasoro than in the $\u1$ \reps .
The missing states are called {\it null states} of the 
Virasoro \rep, which is said to be {\it degenerate} \cite{gins}.

The $\su2$ algebra has two representations of ``spin''
$\sigma=0,1/2$, whose characters are \cite{ciz}:
\barr
\chi^{\su2}_{\sigma=0} & = & 
{1\over \eta(q)^3}\ \sum_{n\in {\bf Z}}\ \left(6n+1\right)\ 
q^{(6n+1)^2/12}\ =\  
\sum_{n\in {\bf Z}}\ q^{n^2}/\eta(q)\ , \nl
\chi^{\su2}_{\sigma=1/2} & = & 
{1\over \eta(q)^3}\ \sum_{n\in {\bf Z}}\ \left(6n+2\right)\ 
q^{(6n+2)^2/12}\ =\  
\sum_{n\in {\bf Z}}\ q^{(2n+1)^2/4}/\eta(q)\ .
\label{char2}\earr
{}From these characters, we obtain the decompositions\footnote{
A similar discussion is found in Ref.\cite{vvvd}.} of \reps
\barr
\{\sigma=0\}_{\su2} &=&\sum_{n \in {\bf Z}\ {\rm even}}\ \{Q=n\}_{\u1} \ =\ 
\sum_{k=0}^\infty \ \left(2s+1\right) \{s=k\}_{\rm Vir}\ ,\nl
\left\{\sigma={1\over 2}\right\}_{\su2} &=&
\sum_{n \in {\bf Z}\ {\rm odd}}\ \{Q=n\}_{\u1} \ =\ \sum_{k=0}^\infty 
\ \left(2s+1\right) \left\{s=k+{1\over2}\right\}_{\rm Vir}\ .
\label{deco}\earr
These results show that:
\begin{itemize}
\item  
The $\su2$ ``spin'' $\sigma$ is only a spin parity.
The $\u1$ charge $Q$ is not a spin quantum number because it is additive.
The Virasoro weight $s=n/2$ composes instead as a good spin quantum number.
\item 
The non-minimal $\u1\otimes\su2$ theory has one \rep for each 
$\sigma=0, 1/2$ value (see the next section);
equation (\ref{deco}) shows that this theory contains the correct number of 
$\winf$ \reps for making spin multiplets.
Therefore,  the non-minimal theory realizes an incompressible quantum fluid
with full $SU(2)$ symmetry.
\item 
The minimal $\winf$ theory contains just one
\rep for each value of the spin $s$ and does not have the full 
$SU(2)$ symmetry, although each $\winf$ excitation has associated a spin 
weight; this is a ``hidden'' symmetry.
For $s=1$ in particular, there is only $J_+$ out of the three currents
$\{J_+,J_0,J_-\}$. These features extend to generic $m$: the partial
reduction of the $SU(m)$ symmetry in the ${\cal W}_m$ theories can be 
understood in the framework of the Hamiltonian reduction of 
Ref.\cite{hamred}.
\end{itemize}

Given these algebraic properties, the physical motivations
for preferring either the minimal or the non-minimal $\winf$ theories 
should come from the understanding of the $SU(m)$ symmetry
of the Jain plateaus in the neutral edge spectrum.
At present, we do not have sufficient numerical and experimental evidence
in favour of either hidden or manifest symmetry; on the theoretical side, 
a fundamental derivation of the Jain approach is also missing.
Hereafter, we collect the pieces of evidence we are aware of:
\begin{itemize}
\item 
The minimal models (hidden $SU(m)$ symmetry) realize the simplest 
consistent theory allowed by the $\winf$ symmetry and its fusion rules.
There are no additional conserved charges besides the electric,
a rather economic and, thus, appealing feature. 
The reduced number of excitations
with respect to the generic hierarchical Abelian theories could be
related to the stronger stability of the Jain plateaus \cite{ctz5}.
\item 
On the other hand, the $\u1\otimes\suem$ symmetry could be motivated as
follows. In the Jain approach, the $\nu=m/(ms\pm 1)$ and the $\nu=m$
plateaus are qualitatively similar \cite{jain}. 
One can consider the naive model of $N$ {\it non-interacting} electrons 
filling $m$ Landau levels on a disk: the (particle-hole) edge excitations 
of each filled level gives rise, for $N\to\infty$, to an independent copy 
of the $\nu=1$ edge, which is described by a Weyl fermion $c=1$ theory
\cite{cdtz1}. These $m$ Weyl fermions have naturally associated the 
$\u1\otimes\suem$ symmetry of unitary rotations in spinor space.
\item 
This hypothesis of many independent and equivalent edges is
supported by a numerical simulation at $\nu=2/5$ $(m=2)$ \cite{twoll}. 
The spectrum of the microscopic Hamiltonian for few electrons was found in a
two Landau-level Hilbert space, as a function of the cyclotron energy 
$\omega_c$.
The incompressible fluid ground state was shown to remain in 
the same universality class all along the range from zero to a large value
$\omega_c$. The vanishing limit can be considered as a model for 
the independent edges, while the infinite limit is closer to the 
experimental setting of large magnetic fields.
\item 
However, these effective $m$ Landau levels are certainly not equivalent 
in the microscopic theory \cite{jj}: there are cyclotron gaps among them, 
which imply a band structure of the quasi-particle excitations; moreover, 
they have different wave functions even for $\omega_c=0$.
We do not yet known an explicit derivation of the edge excitations from a 
realistic microscopic theory of interacting electrons. 
\end{itemize}

Finally, we describe a different experimental situation in which the
$\u1\otimes\suem$ theories and their $SU(2)$ symmetry is fully justified.
This is the spin-singlet quantum Hall effect found at $\nu=2/3$ in samples 
of low electron density \cite{tilt}. Since the same filling fraction is
achieved for smaller magnetic fields, the Zeeman energy
is reduced and the electrons may anti-align.
Experimental evidence was found for a phase transition between
a spin-singlet and the usual spin-polarised ground state at
$\nu=2/3$, when the magnetic field is tilted with respect to the vertical 
of the plane.
According to the Refs.\cite{spinsing}, the new universality class of the 
spin-singlet incompressible fluid is still described by the
$\u1\otimes\su2$ theory (\ref{wspec}) of the spin-polarised
phase, where the two fluids are now interpreted as spin states.
Above the spin-singlet ground state, the edge excitations carry a spin,
which can be naturally identified with the isospin number $s$ of degenerate
$\winf$ \reps. Actually, they occur in the right multiplicity and give 
the elementary, i.e. irreducible, excitations of the incompressible fluid.
Therefore, the degenerate $\winf$ \reps are a useful basis for making 
explicit the $SU(m)$ symmetry of the $\u1\otimes\suem$ theories.

%- 4 -------------------

\section{Partition functions for hierarchical plateaus}
\label{sec4}

\subsection{Diagonal modular invariant for $\u1^{\otimes m}$ theories}

We now focus our attention on minimal and non-minimal $\winf$ theories with 
positive integer central charge $c=m=2,3,\dots$. 
We first consider the $\u1^{\otimes m}$ theories, describing
generic hierarchical plateaus, whose one-edge spectrum (\ref{mabel}) is
parametrised by an integer symmetric $K$ matrix.
We shall now find the corresponding annulus partition function. 
We take a lattice of generic $\u1^{\otimes m}$ representations 
which is closed under the fusion rules. Their weights are, in
$m$-dimensional vector notation, 
\beq
Q \ =\ t^T\ K^{-1} n\ ,\qquad L_0 \ = {1\over 2}\ n^T\ K^{-1}\ n\ ,
\label{mrap}\eeq
where $K$ is a real symmetric positive-definite
matrix and $\vec{t}$ is a real vector of charge units.
We look for partition functions of the form Eq.(\ref{rcftz}) properly
generalised to the multi-component case, which are
solutions of the conditions $(T^2,S,U,V)$ in Eqs. (\ref{tcond}) -
(\ref{vcond}). As in the $c=1$ case, we first consider the $U$ condition by
collecting the sectors with integer-spaced charges in the spectrum
(\ref{mrap}). These are given by
$ \vec{n}=K\ell+ \lambda\ $, $\ \vec{\ell} \in {\bf Z}^m$
provided that $\vec{t}$ has integer components. If the matrix $K$ is also
integer, there is a finite number of $\vec{\lambda}$ values 
(the sectors of the RCFT), belonging to the quotient of the $\vec{n}$ lattice 
by the $\vec{\ell}$ lattice:
\beq 
\vec{\lambda} \in {{\bf Z}^m \over K \ {\bf Z}^m} \ .
\label{mlatt}\eeq 
As in section two,  the $\u1$ characters in each sector sum up into 
$m$-dimensional theta functions (see Eq.(\ref{thetaf})): 
\barr
\chi_{\vec{\lambda}} &=&
{\rm e}^{\disp -{\pi\ t^T K^{-1}t}\ {\left(\I\zeta\right)^2\over\I\tau}}
\times\nl
&\  &\frac{1}{\eta(q)^m}\ \sum_{\vec{\ell}\ \in\ {\bf Z}^m}
{\rm e}^{\disp i2\pi \left\{
\frac{\tau}{2} \left(K\ell+\lambda\right)^T K^{-1}\left(K\ell+\lambda\right)
+\zeta\ t^T \left(\ell + K^{-1}\lambda\right) \right\} } . 
\label{mthetaf}\earr
Their $T^2,S,U,V$ transformations read:
\barr
T^2:\ \chi_{\vec{\lambda}}\left(\tau+2,\zeta\right) & = &
{\rm e}^{\disp\ i2\pi\left(\lambda^TK^{-1}\lambda-{m\over 12} \right) }
\ \chi_{\vec{\lambda}}\left(\tau,\zeta\right)\ ,\\
S:\ \chi_{\vec{\lambda}}\left(-{1\over\tau},-{\zeta\over\tau}\right) & = &
{{\rm e}^{\disp\ i\pi\ t^T K^{-1} t\ \R {\zeta^2\over\tau}}\over\sqrt{\det K}}
\sum_{\vec{\lambda^\prime}\in {\Z^m\over K\Z^m} }\
{\rm e}^{\disp i2\pi\ {\lambda^T K^{-1}\lambda^\prime} }
\chi_{\vec{\lambda^\prime}}\left(\tau,\zeta\right),\\
U:\ \chi_{\vec\lambda}\left(\tau,\zeta+1\right) & = &
{\rm e}^{\disp\ i2\pi\ {t^T K^{-1}\lambda} } \
\chi_{\vec\lambda}\left(\tau,\zeta\right)\ ,\\
V:\ \chi_{\vec\lambda}\left(\tau,\zeta+\tau\right) & = &
{\rm e}^{\disp -i{2\pi\ t^T K^{-1}t }\left(\R\zeta +\R{\tau\over 2}\right)}
\chi_{{\vec\lambda}+\vec{t}}\left(\tau,\zeta\right)\ .
\label{mchitr}\earr
Therefore, they carry a finite-dimensional unitary projective \rep of the 
modular group $\Gamma (2)$.
According to the discussion in section two, the dimension of this 
\rep gives the Wen topological order (see Eq.(\ref{topan})),
\beq
dim {\cal H}(\Sigma_1)= |\det K| \ .   
\eeq

Some physical conditions further constrain the parameters $\vec{t}$ and $K$.
By extending the $c=1$ analysis of section two, one requires
$m$ electron states in the chiral spectrum, having unit charge and odd 
statistics, and integer statistics relative to all the states.
These conditions imply, in a suitable basis, 
that $t_i=1\ $ and $K_{ii}$ is odd for $i=1,\dots, m$.
Moreover, the Hall current $\ \nu=t^TK^{-1}t\ $ is obtained from the 
transport of the minimal fractional charge $\vec{\lambda}=\vec{t}$ 
between the two edges.
The $U$ invariance of the partition function, written
as a sesquilinear form of the characters (\ref{rcftz}), implies the equation
$\ t^T K^{-1} \left(\lambda-\bar\lambda\right)\ \in {\bf Z}$
for the left and right weights. Its solutions depend on the specific 
form of $K$; here, we shall only discuss the general {\it diagonal} 
solution, $\vec{\lambda}=\vec{\bar{\lambda}}$ .
This is also a solution of the other $(T^2,S,V)$ conditions.
The modular invariant partition function is thus a diagonal quadratic form
of the generalised characters (\ref{mthetaf}),
\beq
Z\ = \ \sum_{\vec{\lambda}\in \Z^m/K\Z^m}\ \chi_{\vec{\lambda}}\ 
\bar\chi_{\vec{\lambda}} \ .
\label{zem}\eeq
This partition function contains the chiral spectrum (\ref{mabel}) 
for each edge,  in the case of positive-definite\footnote{ 
The case of mixed chiralities will be described later.} $K$.
Moreover, it describes pairs of left-right excitations with integer total
charge. Note that it is again an odd variant of the many-component boson 
compactified on a rational torus \cite{gins}. Some results on these partition 
functions were also found in Refs.\cite{esko}\cite{read}.

\subsection{Diagonal invariant for the $\u1\otimes\suem$ theories}

Next, we describe more specific results for the non-minimal and
minimal $\winf$ models describing the Jain plateaus.
First we rederive the diagonal modular invariant (\ref{zem})
in a simpler two-dimensional basis with explicit $\suem$ symmetry.
In this basis, we make contact with Itzykson's analysis of
$\suem$ modular invariants \cite{itz}: although our setting
is rather different, this analysis will be used to find many
non-trivial $\u1\otimes\suem$ invariants, which define new edge theories.

The $\suem$ \reps and characters were described in Ref.\cite{itz}: 
there are $m$ highest weight representations,
corresponding to completely antisymmetric tensor  \reps of the 
$SU(m)$ Lie algebra,
which can be labelled by $\alpha=1,\dots,m$. This is an additive
quantum number modulo $m$, the so-called $m$-ality. 
Therefore, the $\suem$ fusion rules are isomorphic to the ${\bf Z}_m$ group.
We shall assume that the $\suem$ excitations do not carry directly
a charge quantum number; thus, we shall need
their characters for $\zeta=0$, which satisfy
\beq
\chi^{\suem}_\alpha(\tau,0)= \chi^{\suem}_{m-\alpha}(\tau,0) = 
\chi^{\suem}_{m+\alpha}(\tau,0)\ ; 
\label{csuem}\eeq
the Virasoro dimension of the highest weight states  is
\beq
L_0={\alpha(m-\alpha)\over 2m}\ ,\qquad \alpha=0,\dots,m-1\ .
\eeq
As shown for the $m=2$ case in section three (Eq.(\ref{deco})), these 
characters resum an infinity of degenerate $\winf$ \reps , which have 
dimensions $L_0 + {\bf Z}$, and $SU(m)$ weights of the same $m$-ality $\alpha$.
The modular transformations are \cite{itz}
\barr
T^2 \ :\ \chi^{\suem}_\alpha\left(\tau+2\right) &=&
{\rm e}^{\disp i2\pi\left({\alpha(m-\alpha)\over m}-{m\over 12}\right)}
\ \chi^{\suem}_\alpha\left(\tau\right)\ ,\nl
S:\ \chi^{\suem}_\alpha \left(-{1\over\tau}\right) & = &
{1\over\sqrt{m}}\ \sum_{\alpha^\prime=1}^m\
{\rm e}^{\disp -i2\pi{\alpha\alpha^\prime\over m} } \
\chi^{\suem}_{\alpha^\prime}\left(\tau\right)\ ,
\label{chim}\earr
while $U,V$ do not act on neutral states.

Furthermore, the $\u1$ factor contained in the $\u1\otimes\suem$ theories 
can be described by the finite set of ${\bf Z}_p$ characters (\ref{thetaf}) 
of section two, with free parameter $p\in {\bf Z}$. They will be denoted as
$\chi^{\u1}_\lambda(\tau,\zeta)$, with $\ \lambda=1,\dots,p$.
In summary, the modular invariance problem has been casted
into a two-dimensional basis, where the $S$ transformation
acts as the ${\bf Z}_p\times {\bf Z}_m$ discrete Fourier
transform, plus additional physical conditions.
Clearly, we should discard the solutions with decoupled 
$\u1$ and $\suem$ sectors, because these are simple superpositions
of a Laughlin fluid and an independent neutral fluid.

The Jain spectrum (\ref{wspec}) possesses the $\suem$ symmetry and can be
rewritten in this basis. Upon substitution of
\barr
n_1 &=& l+\sum_{i=2}^{m} k_i \pm \a\ ,\nl
n_i &=& l - k_i\ ,\ i=2,\dots,m\ ,
\label{chba}\earr
where $l,k_i\in{\bf Z}$ and $\alpha $ is the $\suem$ weight,
the Jain spectrum becomes:
%\vbox{
\barr
\nu={m\over ms \pm 1}\quad :
\ Q & = & \frac{ ml \pm \alpha}{ms \pm 1}\ \nl
J &=& {\left(ml \pm\alpha\right)^2 \over 2 m(ms \pm 1)} \pm 
\frac{\alpha(m-\alpha)}{2m} +\ r\ , \quad r \in{\bf Z}\ .
\label{usspec}\earr
%}
Consider now these formulas with the upper signs, the other choice
will be discussed later.
One recognises the $\u1$ and $\suem$ contributions to $L_0$ ($J$),
which identify $p=m\hat p$, $\hat p=(ms+ 1)$ and 
$\lambda=ml +\alpha\ $ mod $(m\hat p)$; note that $\hat p= 1$ mod $m$ and
$\lambda=\alpha$ mod $m$, and that $\hat p, m$ are coprime numbers 
$(\hat p,m)=1$.
Consider the $m$-term linear combinations of the $(m^2\hat p)$
tensor characters
\beq
\theta_\lambda\left(\tau,\zeta\right) = \sum_{\alpha=1}^m\  
\chi^{\u1}_{\lambda+\alpha\hat p}\left(\tau,m\zeta\right)\ 
\chi^{\suem}_{\lambda +\alpha\hat p\ {\rm mod}\ m}\left(\tau,0\right)\ , 
\qquad \lambda=1,\dots, m\hat p\ .
\label{ddiag}\eeq
They satisfy $\theta_{\lambda+\hat p}=\theta_\lambda$, due to 
$\hat p=1$ mod $m$;
thus, there are $\hat p$ independent ones, which can be chosen to be
$\theta_{ma}$, $a=1,\dots,\hat p$.
One can check that they carry a representation of the modular
transformations $(T^2,S,U,V)$, in particular
$S_{ab}\propto\exp\left(i2\pi\ mab/\hat p\right)/\sqrt{\hat p}\ $;
its dimension matches the Wen topological order $det(1+ sC)=\hat p$. 
The unique solution of the $U$ condition in the basis
$\theta_\lambda$ is given by the diagonal pairing of left and right 
characters,
\beq
Z=\sum_{a=1}^{\hat p}\ \theta_{ma}\ \bar{\theta}_{ma}=
\sum_{a=1}^{\hat p}\  
\left(\sum_{\a=1}^m\ \chi^{\u1}_{ma+\a\hat p}\ \chi^{\suem}_\a \right)
\left(\sum_{\b=1}^m\ \bar\chi^{\u1}_{ma+\b\hat p}
       \ \bar\chi^{\suem}_\b \right)\ .
\label{dsuem}\eeq
One can check explicitly that the characters $\theta_{ma}$ 
are equal to the $\chi_{\vec{\lambda}}$ in (\ref{mthetaf}) for 
$\ K=1+sC\ $, once the corresponding $\u1$ charges are identified; 
the partition function (\ref{dsuem}) is equal to (\ref{zem}), and
the spectrum of charges and fractional spins reproduces 
(\ref{wspec}), with upper signs, for each chirality. 
The $V$ transformation is
\beq
V \ :\ \theta_{ma}\left(\tau,\zeta+1\right) =
{\rm e}^{\disp -i2\pi {m\over \hat p}
\left(\R \zeta +\R {\tau\over 2}\right)}
\ \theta_{ma+m}\left(\tau,\zeta \right)\ .
\label{stillv}\eeq
This shows that the minimal transport of charge between the two edges is 
$m$ times the elementary fractional charge, which is the smallest spectral
flow at constant $\alpha$ among the states contained in (\ref{dsuem}).
The Hall current is thus $\nu=m/\hat p$.
 
Finally, it remains to check the existence of {\it one} electron 
excitation for each edge, which is described by the states $(a,\alpha)$ 
in $Z$ (\ref{dsuem}) satisfying
\barr
Q&=&{m(a+k\hat p)+\alpha\hat p \over \hat p}=1\ ,\qquad k\in {\bf Z}\ ,\nl 
2J&=&{\left( m(a+k\hat p)+\alpha\hat p\right)^2 \over m\hat p} +
{\a (m-\a) \over m} = 1\quad {\rm mod}\ \ 2\ ,
\label{elcond}\earr
with $\hat p=1+ms$. This is solved by $\alpha=1$ and 
$a=0$, provided that $s$ is even.
Moreover, the condition (\ref{estat}) is also satisfied, such that the
electron ($\alpha=1$) is local, i.e. it has integer monodromy with all
excitations. 
We thus found the partition functions (\ref{dsuem}) of the $\u1\otimes\suem$
theories at the Jain plateaus; they describe rational conformal field 
theories with an extension of the affine algebra $\u1\otimes\suem$ and 
fusion algebra ${\bf Z}_{\hat p}$.

We now find the partition function for the case of mixed chiralities
in the one-edge spectrum (\ref{wspec}), at $\nu=m/(ms -1)$.
The fractional spin is not positive-definite and cannot be
identified with the Virasoro dimension. 
Moreover, the spectrum is not decomposable into chiral and anti-chiral
independent parts: $K\neq K^R-K^L$, with $K^L,K^R$ positive definite
in the notation of (\ref{mrap}). 
Nevertheless, a simple modification of the characters (\ref{ddiag}) switches
the chirality of the neutral excitations,
\beq
\theta^{(-)}_\lambda = \sum_{\a =1}^m\  
\chi^{\u1}_{\lambda+\a\hat p}\ 
\bar\chi^{\suem}_{\lambda +\a\hat p\ {\rm mod}\ m}\ , 
\qquad \lambda=1,\dots, m\hat p=m(ms-1)\ .
\label{antic}\eeq
This gives again a representation of the modular group for $\hat p=ms-1>0$. 
The $U$ condition implies the partition function 
\beq
Z=\sum_{a=1}^{\hat p}\  
\left(\sum_{\a =1}^m\ \chi^{\u1}_{ma+\a\hat p}\ \bar\chi^{\suem}_\a \right)
\left(\sum_{\b =1}^m\ \bar\chi^{\u1}_{ma+\b\hat p}\ \chi^{\suem}_\b \right)\ .
\label{dcsuem}\eeq
which satisfies all the conditions and reproduce the spectrum 
(\ref{wspec}) for $\nu=m/(ms-1)$.
Note that the spectrum of $\left(L^R_0+L^L_0 \right)$ is still positive 
definite and ensures the convergence of the sum. 
This partition function combines representations of the two chiralities
in a non-trivial way, thus it difficult to assign excitations to
either the inner or the outer edge. From the physical point of
view, we are already used to non-local effects in the 
incompressible fluid, like fractional statistics. Moreover,
this spectrum has been partially confirmed by experiments.
{}From the theoretical point of view, this was not easily accounted for
by the chiral theories of one edge, because it is not left-right 
decomposable \cite{zee}\cite{ctz5}.
Instead, it is naturally described by annulus partitions functions
which are not left-right diagonal.

\subsection{Minimal $\winf$ models}

As discussed in section three, the partition functions 
(\ref{dsuem}),(\ref{dcsuem}) can be decomposed into degenerate $\winf$ 
representations, owing to the algebra inclusions (\ref{inclu}).
For $m=2$, we can use the relations among the characters (\ref{char2})
and find that the $\winf$ \rep of spin $\ell$ occurs with multiplicity 
$(2\ell +1)$. Therefore, these partition functions describe
non-minimal $\winf$ theories which possess a full $SU(2)$ symmetry.

Actually, these seem to be the {\it unique} RCFT with $\winf$ symmetry.
Indeed, the $(m=2)$ $\winf$ characters 
yield a linear finite-dimensional \rep of the $S$ transformation
only if they are summed up with multiplicity $(2\ell +1)$. 
Here is the argument. The non-trivial part of the character is the 
${\cal W}_2$, i.e. Virasoro, part (\ref{chiu}), which we rewrite
\beq
\chi^{\rm Vir}_d={1\over \eta} \left(
q^{(d-1)^2/4} - q^{(d+1)^2/4} \right)
= - \chi^{\rm Vir}_{-d}\ ,\qquad d=2\ell+1 \ge 1 \ .
\label{chivir}\eeq
Consider the sum of these characters with generic multiplicities $N_d$, 
extended odd for $d<0$ for convenience, 
\beq
\chi[N] =\sum_{d=1}^\infty \ N_d \chi^{\rm Vir}_d =
\sum_{k\in {\bf Z}}\ N_k\ {q^{(k-1)^2/4} \over \eta} \ ,
\qquad N_d =- N_{-d} > 0 \ {\rm for}\ d >0 \ .
\label{gensum}\eeq
It is rather easy to compute the $S$ transformation of $\chi[N]$
by using the Poisson formula (appendix A),
and check that it reproduces itself, up to phases, only for the
multiplicity $N_d=d$ leading to the $\widehat{SU(2)}_1$ characters
(\ref{char2}).

Therefore, we cannot form a modular invariant partition function 
of the RCFT type (\ref{rcftz}) for the $\winf$ minimal models \cite{ctz5},
which contain $\winf$ \reps with multiplicities $N_d=1$, for $d\ge 1$:
the $\winf$ minimal models are {\it non-rational} conformal field 
theories. 

One can tentatively disregard the $S$ modular invariance and
build minimal $\winf$ partition functions, subjected to the remaining 
conditions $(T^2,U,V)$ and the existence of the electron state. 
However, it turns out that these building criteria are not sufficient,
because they allow a very large class of consistent models. 
Their fusion rules can close on a subset
of degenerate $\winf$ \reps with $SU(m)$ weights of given $m$-alities 
$\alpha=n\delta$, $n=1,\dots,m/\delta$, where $\delta$ divides $m$ and 
satisfies $(\delta,m/\delta)=1$. 
These solutions are discussed in Appendix B and were not considered in 
Ref.\cite{ctz5}.
They possess Hall conductivities which span almost 
any fraction $\nu=n/d$ and are thus phenomenologically unrealistic.
Following the discussion of section three, we conclude that:

i) If the minimal $\winf$ theories are realized at the Jain plateaus 
(no full $SU(m)$ symmetry of the edge excitations), then another building 
requirement, replacing modular invariance, is necessary to construct 
their partition functions. These non-rational conformal theories are
not well understood in the literature and are not further discussed in
this paper.

ii) If the non-minimal, $\u1\otimes\suem$ models are realized (full
$SU(m)$ symmetry), then their partition functions can be found in 
this paper.

\subsection{Non-diagonal $\u1\otimes\suem$ invariants}

All modular invariant partition functions can be divided 
in two classes, according to the properties of
the associated fusion algebras \cite{mose}: 
\begin{itemize}
\item
There are invariants describing extensions of the symmetry algebra by 
some dimension-one field in the theory. For example, the $\theta_\lambda$ 
in (\ref{stillv}) are characters of an extended algebra 
\reps which finitely decompose into those of $\u1\otimes\suem$. 
The partition function (\ref{dsuem}) is diagonal in the extended basis
$(\theta_\lambda)$, but actually non-diagonal in the original basis
$\left(\chi^{\u1}_\lambda\ \chi^{\suem}_{\a}\right)$.
\item
Once the algebra is maximally extended, 
the non-diagonal invariants are necessarily of the form
$Z=\sum\ \chi_\lambda\ \bar\chi_{\pi(\lambda)}$,
where $\lambda\to\pi(\lambda)$ is a permutation of $\lambda$ and an 
automorphism of the fusion rules, $N_{ijk}=N_{\pi(i),\pi(j),\pi(k)}$.
Indeed, these $Z$ are invariant under the $S$ transformation, 
$[\pi,S]=0$, due to the Verlinde relation between fusion rules and the
$S$ transformation (\ref{opes}); moreover, $[\pi,T]=0$ by hypothesis.
\end{itemize}
The modular invariance problem in this paper is slightly non-standard,
because there are additional conditions $U,V$ concerning the $\u1$ charge, 
and some physical conditions. Actually, we shall find that 
the extensions of the $\u1$ algebra are trivial, but are
possible for the $\u1\otimes\suem$ algebra; moreover,
the automorphisms involving the $U(1)$ charge are forbidden by the 
$U$ condition, while they are possible for the neutral $\suem$ weight
in a limited sense. 

In order to proceed, we need to be more specific about the general
form of the invariants $Z=\chi\cdot{\cal N}\cdot\bar\chi$, 
characterised by matrices ${\cal N}_{\lambda\bar\lambda}$, 
which commute with the modular transformations $(T^2,S,U,V)$.
Let us first discuss the $\u1$ theory of section two and 
prove that the unique solution is given by the diagonal partition 
function (\ref{zedone}), namely 
${\cal N}_{\lambda\bar\lambda}=\delta_{\lambda,\bar\lambda}$,
where $\lambda$ labels the ${\bf Z}_p$ extended $\u1$ characters 
in (\ref{thetaf}).
The ${\cal N}$ matrices commuting with the $c=1$ transformations
$(T^2,S)$ in (\ref{chitr}) can be found by generalising the method
of Refs.\cite{ciz}\cite{itz} - the present case only differs by some 
factors of two. The commutation with $T^2$ implies, 
\beq
{\cal N}_{\lambda,\bar\lambda}\neq 0\ \ {\rm for}\ \ 
\lambda^2=\bar\lambda^2 \qquad {\rm mod} \ p\ .
\label{t2comm}\eeq
Following Ref.\cite{ciz} closely, we consider a decomposition into factors 
$p=rs$, where $\delta=(r,s)$ is their common factor, and rewrite 
(\ref{t2comm})
as the system $\lambda+\bar\lambda =r\rho$, $\lambda-\bar\lambda = s\sigma$.
The solutions are pairs $(\lambda,\bar\lambda)$ multiple of $\delta$
$(\delta/2)$ for $\delta$ odd (even), respectively, which can be
written as follows. 
Consider first $\delta$ odd. Define $\omega$ by the system
\beq
\left\{ \begin{array}{rllll}
1= {\disp r\over\disp \delta} R- {\disp s\over\disp \delta}S\ , 
& \quad &  \left(R,S\right) &{\rm mod} & 
\left({\disp s\over\disp \delta},{\disp r\over\disp \delta}\right)\ ,\\
& & & & \\
\omega= {\disp r\over\disp \delta} R + {\disp s\over\disp \delta}S\ , 
&\quad & \omega &{\rm mod} & 2 p/\delta^2\ , \end{array} \right. 
\label{perid}\eeq
which satisfies $\omega^2=1$ mod $4p/\delta^2$.
We actually need $\omega$ mod $p/\delta^2$.
The solutions $\lambda,\bar\lambda$ multiples of $\delta$, can be shown to
satisfy
\beq
\lambda =\omega\ \bar\lambda +\xi {p\over\delta} \ \ \ {\rm mod}\ p\ ,
\quad \xi=1,\dots,\delta, \qquad {\rm for}\ \ \delta\vert\lambda\ ,
\ \delta\ \vert\bar\lambda\ ,\quad (\delta\ {\rm odd}).
\label{t2sol}\eeq
Moreover, these give the following solution to the $S$ condition
${\cal N}S=S{\cal N}$:
\beq
{\cal N}_{\lambda\bar\lambda}=\left\{ \begin{array}{ll}
\disp \sum_{\xi=1}^\delta\ 
\delta^{(p)}_{\lambda,\omega\bar\lambda +\xi p/\delta}\ ,\quad
& {\rm if}\ \ \delta \vert\lambda\ , \delta\vert\bar\lambda \ , \\
0 \ , & \ {\rm otherwise} \ .\end{array} \right.
\label{ssol}\eeq

If $\delta$ is even, define $\delta=2\delta^{\prime}$ and 
$p=4\delta^{\prime 2}k$. Define $\omega$ mod $p/2\delta^{\prime 2}$ 
by Eq.(\ref{perid}); we need both 
$\omega=\omega_1$ and $\omega_2=\omega +p/2\delta^{\prime 2}$ mod 
$p/\delta^{\prime 2}$.
The solutions to both $T^2$ and $S$ are again of the form
(\ref{ssol}) with $\delta\to\delta^\prime$, non-vanishing for the indices 
$(\lambda,\bar\lambda)$ multiples of $\delta^\prime$, 
which satisfy (\ref{t2sol}) with $\omega\to\omega_i$, for $i=1,2$.
In conclusion, we found one (two) solutions for $\delta$ odd (even),
respectively. One can probably extends the methods of Ref.\cite{ciz} for
proving that these give the general solution.

These solutions show explicitly the two general classes
of invariants discussed before: for $\delta\neq 1$ 
and $\omega=\pm 1$, the partition functions (\ref{ssol}) are diagonal sums
of the generalised characters
\beq
\Theta_{\a^\prime}=\sum_{\xi=1}^\delta 
\chi_{\delta\a^\prime+\xi p/\delta} \ , \qquad \a^\prime=1,\dots,p/\delta^2\ ,
\label{extchi}\eeq
for \reps of an extended chiral algebra, which transform by 
$S_{\a^\prime}^{\b^\prime}=$ \\
$\exp \left(-2i\pi \a^\prime\b^\prime\delta^2/p \right)$.
Moreover, solutions (\ref{ssol}) with $\omega\neq\pm 1$ and $\delta =1$
pair non-trivially the left and right algebras by an automorphism
of the fusion rules.

In the $\u1$ theory, these solutions do not pass 
the other conditions or are trivial. Actually, the extended solutions are 
trivial, because the sums (\ref{extchi}) of $\u1$ characters 
(\ref{thetaf}) satisfy the identity
$\Theta_{n}(p)\propto\chi^{\u1}_n(p/\delta^2)$, i.e. they reproduce
themselves at another value $p\to p/\delta^2$, which is considered anyhow.
Moreover, the automorphisms $\omega$ are killed by the $U$ 
condition (\ref{ucond}), which implies $\lambda=\bar\lambda$ mod $p$, i.e.
$\omega=1$ mod $p/\delta^2$ for
$\delta$ odd (respectively, mod $4p/\delta^2$ for $\delta$ even). 

We now come back to the discussion of non-trivial invariants for
$\u1\otimes\suem$. This problem is two-dimensional: 
the conditions $S,T^2$ for the pure $\u1$ part were discussed before, 
while those of the $\suem$ part have the same solutions (\ref{perid}) -
(\ref{ssol}), with $\lambda$ replaced by $\alpha$ and $p$ by $m$.
Actually, $S$ is a ${\bf Z}_m$ discrete Fourier 
transform and $T^2$ implies $\alpha^2 =\bar\alpha^2 $ mod $m$, 
according to Eq.(\ref{chim}).
Therefore, we know the general solutions for each of the one-dimensional
sub-problems.
Here, we shall not reach a complete solution of the two-dimensional problem, 
but present some two-dimensional coupled invariants which are suggested by
the one-dimensional solutions.

Let us first consider the new invariants associated to extensions 
of the $\suem$ algebra for $m=\delta^2 m^\prime$. For 
$\hat p=1+m^\prime s$, the following invariants generalise (\ref{dsuem}):
\barr
Z &=&\sum_{a=1}^{\hat p}\  
\left(\sum_{\a^\prime=1}^{m^\prime}\ \chi^{\u1}_{m^\prime a+\a^\prime\hat p}\ 
\Theta^{\suem}_{\a^\prime} \right)
\left(\sum_{\b^\prime=1}^{m^\prime}\ 
\bar\chi^{\u1}_{{m^\prime}a+\b^\prime\hat p}\ 
\bar\Theta^{\suem}_{\b^\prime} \right) \ ,\nl
\Theta^{\suem}_{\a^\prime} &=&
 \sum_{\xi=1}^\delta \chi^{\suem}_{\delta(\a^\prime +\xi m^\prime)} 
\qquad\quad (m=\delta^2 m^\prime) \ .
\label{extinv}\earr
These extended characters resum $\suem$ \reps with 
Virasoro dimensions differing by integers or half-integers.
The electron conditions are
sensitive to half-integer spins: in analogy with (\ref{elcond}), the 
electron is described by the extended weights $(a,\a^\prime)$
satisfying
%\vbox{
\barr
Q&=&{m^\prime a\over \hat p} +\a^\prime =1\ ,\quad \hat p=1+m^\prime s\ ,
\qquad \to \ a=0,\ \ \a^\prime=1 \ ,\nl
2J&=& {\left(m^\prime a+\a^\prime\hat p\right)^2\over m^\prime\hat p} + 
{\left(\a^\prime +m^\prime\xi\right)
 \left(\delta m^\prime - \a^\prime -m^\prime\xi\right)\over m^\prime} \nl
&=& s+\delta +m^\prime\xi (\delta -\xi)=1 
\ \ {\rm mod}\ 2\ ,\quad \forall\ \xi=1,\dots,\delta \ .
\label{excond}\earr
%}
For $m=\delta^2 m^\prime$ and $\delta$ odd, the solutions are $s$ even
and any $m^\prime$. These edge theories span again the Jain filling fractions 
$\nu=m^\prime/(m^\prime s+1)$; their neutral spectrum is more involved
than in the theories (\ref{dsuem}), and is described by extensions of the 
$\u1\otimes\widehat{SU(\delta^2m^\prime)}_1$ algebra, with $\delta=1,3,\dots$
Furthermore, the solutions to (\ref{excond}) with $\delta$ even are given by
$s$ odd and $m^\prime$ even; these edge theories exist for 
other filling fractions\footnote{
We include the analogous solution (\ref{dcsuem}) with mixed chiralities.},
\beq
\nu={m^\prime\over m^\prime s \pm 1} \ , \qquad m^\prime \ {\rm even},
\ s\ {\rm odd}\ ,
\label{newfil}\eeq
The first few values of $0<\nu<1$ with small denominator are 
$\nu=2/3,4/5,6/7,\dots,$ $ 2/7,4/13,\dots,2/5,4/11,\dots$.
Note that the condition (\ref{estat}) of locality of the electron state
is fulfilled by all these solutions.  

Next, we discuss the invariants including non-trivial
twists $\omega\neq\pm 1$ (\ref{perid}). 
According to the previous discussion, these cannot occur in the charged 
spectrum, due to the $U$ condition, but can be introduced for the neutral 
weights. 
The twisted versions of the invariants (\ref{extinv}) exist at the same 
values $\hat p=1+m^\prime s$ and $m=\delta^2 m^\prime$:
\beq
Z =\sum_{a=1}^{\hat p} \ \left\vert\   
\sum_{\a^\prime =1}^{m^\prime}\ \chi^{\u1}_{m^\prime a+\a^\prime\hat p}\ 
\sum_{\xi=1}^\delta \ \chi^{\suem}_{\delta(\omega\a^\prime +\xi m^\prime)}
\ \right\vert^2 \ .
\label{twistinv}\eeq
This invariant is again diagonal with respect to the (maximally) extended
chiral algebra, thus $\omega$ is not an automorphism of its fusion rules,
i.e. this is not an example of the automorphism invariants discussed 
before \cite{mose}; actually, the twist is between 
the $\u1$ and $\suem$ weights entering the extended algebra \reps.
Another type of invariant with one $\omega$ twist only, say 
for the right $\suem$ characters, would not satisfy the reality condition 
$\bar Z\neq Z$ and must be discarded.

The values of $\omega$ mod $m^\prime$ for the twisted invariant
(\ref{twistinv}) are found in (\ref{perid}) 
for any factorisation $m=ab$ with $\rho=(a,b)$: for $\rho$ odd, there is
one solution and we identify $\delta=\rho$ in (\ref{twistinv}); 
for $\rho$ even, there are two solutions and $\delta=\rho/2$.
Actually, non-trivial $\omega\neq\pm 1$ mod $m^\prime$ are only possible for 
relatively large $m^\prime$, which correspond to filling fractions with 
large denominators that are not phenomenologically relevant.
The filling fractions take either the Jain values (\ref{wspec})
or the new ones (\ref{newfil}), according to the parity of $s$, which
is determined by the following electron conditions, analogous to
(\ref{excond}):
\beq 
s+ {1-\omega^2 \over m^\prime} +\omega\delta +m^\prime\xi
\left(\delta-\xi \right) = 1 \quad {\rm mod}\ 2\ ,
\qquad \forall\ \xi=1,\dots,m^\prime\ .
\label{twcond}\eeq
For factorisations $m=ab$ with $\rho=(a,b)=\delta$ odd, the solutions are
$s$ even and any $m^\prime$ (independently of $\omega$), leading to the
Jain filling fractions. For factorisations with $\rho=2\delta$ even,
we have $m^\prime=4k$ even and two values $\omega_1,\omega_2$; if $k$ is odd,
equation (\ref{twcond}) reduces to $s+\delta=1$ mod $2$, independent of
$\omega\ $; if $k$ is even, it remains $\omega$ dependent and $s$ takes 
opposite parities for the two $\omega_i$ values.
Note that the electron states obtained by these solutions 
are always local with respect to the other excitations.
By inspection, the lowest non-trivial values of $\omega\neq\pm 1$,
$m^\prime$ and $\nu$ are:
\begin{itemize}
\item 
$\omega= 3$ mod $m=m^\prime=8$, corresponding to filling fractions
$\nu=8/(8s \pm 1)$, $s$ odd, i.e. $\nu=8/9,8/23,8/25, \dots$
\item
$\omega=5$ mod $m=m^\prime=12$ corresponding to $\nu=12/(12s \pm 1)$, 
$s$ even, i.e. $\nu=12/23, 12/25, \dots$.
\end{itemize}

In conclusion, we cannot prove here that the modular invariants found in
this section give a complete solution; there might exist additional
exceptional solutions, i.e. sporadic $(\hat p,m)$ values not falling
into sequences. We also remark that the many non-trivial invariants
(\ref{extinv}), (\ref{twistinv}) possess a rather limited set of 
filling fractions. 
This feature has some phenomenological appeal, as discussed in
the following section.

\begin{figure}
\unitlength=0.8pt
\begin{picture}(500.00,440.00)(-10.00,0.00)
\put(0.00,500.00){\line(1,0){500.00}}
\put(0.00,80.00){\line(1,0){500.00}}
\put(250.00,500.00){\line(0,-1){100.00}}
\put(125.00,500.00){\line(0,-1){100.00}}
\put(83.00,500.00){\line(0,-1){100.00}}
\put(250.00,80.00){\line(0,1){40.00}}
\put(125.00,80.00){\line(0,1){100.00}}
\put(83.00,80.00){\line(0,1){100.00}}
\put(166.00,460.00){\makebox(0,0)[cc]{$\ \ \ \bullet {\bf 1\over 3}$}}
\put(333.00,460.00){\makebox(0,0)[cc]{$\ \ \ \bullet {\bf 2\over 3}$}}
\put(100.00,420.00){\makebox(0,0)[cc]{$\ \ \ \bullet {\bf 1\over 5}$}}
\put(200.00,420.00){\makebox(0,0)[cc]{$\ \ \ \bullet {\bf 2\over 5}$}}
\put(300.00,420.00){\makebox(0,0)[cc]{$\ \ \ \bullet {\bf 3\over 5}$}}
\put(400.00,420.00){\makebox(0,0)[cc]{$\ \ \ \bullet {\it4\over 5}$}}
\put(71.00,380.00){\makebox(0,0)[cc]{$\ \ \ \circ    {\bf 1\over 7}$}}
\put(143.00,380.00){\makebox(0,0)[cc]{$\ \ \ \bullet {\bf 2\over 7}$}}
\put(214.00,380.00){\makebox(0,0)[cc]{$\ \ \ \bullet {\bf 3\over 7}$}}
\put(286.00,380.00){\makebox(0,0)[cc]{$\ \ \ \bullet {\bf 4\over 7}$}}
\put(357.00,380.00){\makebox(0,0)[cc]{$\ \ \ \bullet {\it 5\over 7}$}}
\put(111.00,340.00){\makebox(0,0)[cc]{$\ \ \ \bullet {\bf 2\over 9}$}}
\put(222.00,340.00){\makebox(0,0)[cc]{$\ \ \ \bullet {\bf 4\over 9}$}}
\put(278.00,340.00){\makebox(0,0)[cc]{$\ \ \ \bullet {\bf 5\over 9}$}}
\put(91.00,300.00){\makebox(0,0)[cc]{$\ \ \ \circ    {\bf 2\over 11}$}}
\put(136.00,300.00){\makebox(0,0)[cc]{$\ \ \ \bullet {\bf 3\over 11}$}}
\put(182.00,300.00){\makebox(0,0)[cc]{$\ \ \ \circ   {\it 4\over 11}$}}
\put(227.00,300.00){\makebox(0,0)[cc]{$\ \ \ \bullet {\bf 5\over 11}$}}
\put(273.00,300.00){\makebox(0,0)[cc]{$\ \ \ \bullet {\bf 6\over 11}$}}
\put(318.00,300.00){\makebox(0,0)[cc]{$\ \ \ \circ   {\it 7\over 11}$}}
\put(367.00,300.00){\makebox(0,0)[cc]{$\ \ \ \circ   {\it 8\over 11}$}}
\put(115.00,260.00){\makebox(0,0)[cc]{$\ \ \ \circ   {\bf 3\over 13}$}}
\put(154.00,260.00){\makebox(0,0)[cc]{$\ \ \ \circ   {\it 4\over 13}$}}
\put(231.00,260.00){\makebox(0,0)[cc]{$\ \ \ \bullet {\bf 6\over 13}$}}
\put(269.00,260.00){\makebox(0,0)[cc]{$\ \ \ \bullet {\bf 7\over 13}$}}
\put(308.00,260.00){\makebox(0,0)[cc]{$\ \ \ \circ   {\it 8\over 13}$}}
\put(346.00,260.00){\makebox(0,0)[cc]{$\ \ \ \circ   {\it 9\over 13}$}}
\put(133.00,220.00){\makebox(0,0)[cc]{$\ \ \ \circ   {\bf 4\over 15}$}}
\put(233.00,220.00){\makebox(0,0)[cc]{$\ \ \ \circ   {\bf 7\over 15}$}}
\put(267.00,220.00){\makebox(0,0)[cc]{$\ \ \ \circ   {\bf 8\over 15}$}}
\put(235.00,180.00){\makebox(0,0)[cc]{$\ \ \ \circ   {\bf 8\over 17}$}}
\put(265.00,180.00){\makebox(0,0)[cc]{$\ \ \ \circ   {\bf 9\over 17}$}}
\put(237.00,140.00){\makebox(0,0)[cc]{$\ \ \ \circ   {\bf 9\over 19}$}}
\put(83.00,60.00){\makebox(0,0)[cc]{${\bf 1\over 6}$}}
\put(125.00,60.00){\makebox(0,0)[cc]{${\bf 1\over 4}$}}
\put(250.00,60.00){\makebox(0,0)[cc]{${\bf 1\over 2}$}}
\put(0.00,60.00){\makebox(0,0)[cc]{${\bf 0}$}}
\put(500.00,60.00){\makebox(0,0)[rc]{{\Large$\nu$}$ \qquad {\bf 1}$}}
\end{picture}
\caption{
Experimentally observed plateaus: their Hall conductivity 
$\sigma_H=(e^2/h)\nu$ is displayed in units of $(e^2/h)$.
The marks $\ (\bullet)\ $ denote stable (i.e. large) plateaus, which have 
been seen in several experiments; the marks $\ (\circ)\ $ 
denote less developed plateaus and plateaus found in one
experiment only. Plateaus belonging to the Jain main series have their
filling fraction in {\bf bold} type, the other ones are in {\it italics}.
Coexisting fluids at the same filling fraction
have been found at $\nu=2/3,2/5,3/5,5/7$.
}
\end{figure}
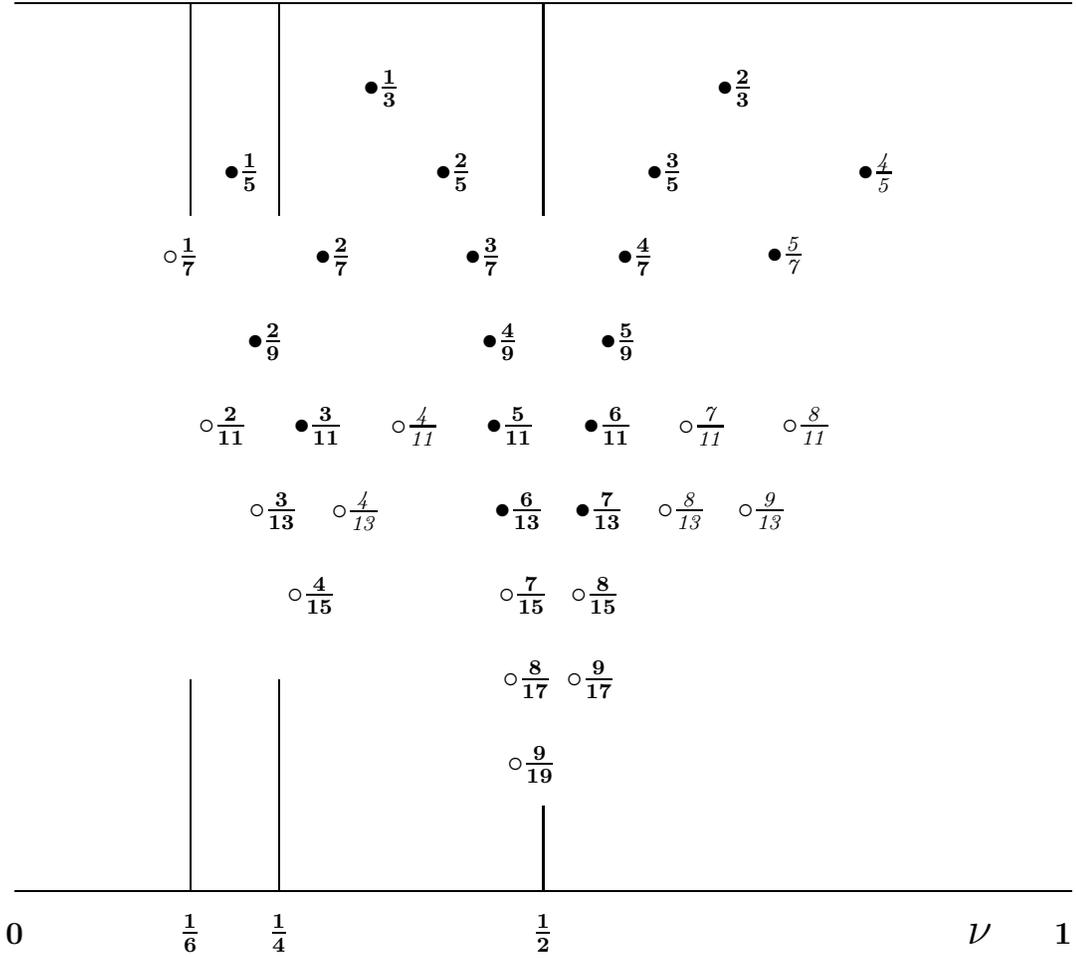

%- 4.3 --------------

\subsection{Comparison with the experiments: higher order hierarchical
pattern}

The experimental values of the filling fraction $0<\nu=n/d< 1$ are
described in figure one, where they are ordered on the vertical axis
according to the value of their denominator \cite{survey}. 
Actually, within the mean field theory \cite{mfth}, the size of the gap of 
the incompressible fluid is of order $O(1/d)$, thus stability decreases 
towards the bottom of the figure. These fractions are usually determined by 
the minima of the longitudinal resistivity and the intersect of
the plateau in the Hall resistivity with the classical curve 
$\rho\propto 1/B$. This is not a very accurate measure, because large,
stable plateaus might hide closer, weaker ones, as for example
near $\nu=1$. Moreover, the data are now a few years old,
because the recent experiments have focussed on the dynamical properties 
of the better understood Jain plateaus.
The fractions denoted by an open dot are weak or may have been found 
under special conditions in one experiment only.
Clearly, more than one edge theory can exist for the same filling
fraction, and a phase transition among them may occur as some 
parameter is tuned at fixed ratio of density to magnetic field.
Transitions between different incompressible fluids 
have been observed at $\nu=2/3,2/5,3/5,5/7$ \cite{tilt}; more might be found as
the engineering of samples improves.

The points belonging to the Jain series (\ref{usspec}) 
$\nu=m/(ms \pm1)$, with $s=2,4,6$ are well established 
and their spectra of $\u1\otimes\suem$ edge excitations 
have been confirmed in part \cite{tunn}\cite{milli}. These are denoted by 
filling fractions in bold type. Their annulus partition functions is given by 
equations (\ref{dsuem}) and (\ref{dcsuem}). 

A very interesting problem is to identify the pattern of the
remaining fractions, which are written in italics in figure one.
Previous approaches were based on the iteration of the Jain
hierarchical wave-function construction \cite{jain}, or by the classification
of $\u1^{\otimes m}$ edge theories associated to integral lattices 
in low dimensions \cite{zee}. A generic feature of
these approaches is that they predict too many unobserved points
within the same hypotheses which describe the few, observed ones.
Here, we would like to show that these remaining experimental
points can be consistently 
described by the new edge theories obtained from the partition 
functions (\ref{extinv}) associated to extensions of the $\u1\otimes\suem$ 
symmetry algebra. 

We shall also use the phenomenological hypothesis of ``charge conjugation''
of the incompressible fluid, which states that to any fraction
$0<\nu<1/2$ there is a corresponding value $1/2<(1-\nu)<1$, 
which may be weaker \cite{prange}.
A droplet of charged conjugate fluid can be thought as a $(\nu=1)$ fluid 
with a $(\nu)$ fluid of ``holes'' carved in it. Therefore, it
has an additional, decoupled, $\nu=1$ charged edge excitation with chirality 
opposite to the rest.
Let us check this hypothesis. The $\nu=m/(2m - 1)$ Jain plateaus have 
associated
a second edge theory, which is the conjugate of the $\nu=m/(2m+ 1)$ one;
as said before, many theories for the same filling fraction do not cause a 
problem, because the most stable (simplest) is observed in 
generic experimental conditions. The conjugates of the observed points
of the $\nu=m/(4m \pm 1)$ series exist in part: 
$7/9$ does not exist but would be expected, because $2/9$ is stable;
less relevant is the absence of $10/13$ and $11/15$, because 
$3/13, 4/15$ are themselves weak; the same can be said about
the conjugates of the third Jain series, $1/7, 2/11$. 
The remaining filling fractions are beyond the Jain series and occur in 
the conjugate pairs $(4/11, 7/11)$ and $(4/13, 9/13)$ plus $8/13$, which 
does not fit this pattern because $5/13$ does not exist.
In conclusion, the charge-conjugation hypothesis is reasonable, 
with problems in the missing  $7/9$ and the unexplained $8/13$. 

Within this analysis, there are only two conjugate pairs of points
outside the Jain series, $(4/11, 7/11)$ and $(4/13, 9/13)$. 
Although other analyses are not excluded, this is a rather 
economic description of all the observed points.
The non-trivial invariants (\ref{extinv}), associated to extensions of the
$\u1\otimes\suem$ symmetry, exist for the series of fractions (\ref{newfil}); 
besides repetitions of the Jain points an their conjugates, the values of
smaller denominator are $\nu=4/11, 4/13$, which match the analysis of the 
experimental data.  
Clearly, the small number of points does not allow for a definite
identification of the new pattern.
Nevertheless, it is important that our theories do not predict many 
unobserved plateaus.

\section{Conclusions}

In this paper, we have defined and found the partition functions for the
edge excitations of the quantum Hall effect in the annulus geometry.
We have shown the modular invariance and we have applied the
properties of rational conformal theories to this 
physical problem; in particular, the topological order 
(\ref{topan}),(\ref{topto}) has been identified with the dimension
of the representation of the modular group.
In section four, we have shown that the $\u1\otimes\suem$ edge 
theories can describe the plateaus of the Jain series as well as 
those beyond them, which have been characterised by an extended symmetry 
algebra and the non-trivial partition functions (\ref{extinv}). 
Clearly, a complete solution of the modular invariant 
conditions is needed to accomplish this classification program.
We also found that the $\winf$ minimal models are not rational
conformal field theories.

The partition function and the related RCFT properties are also
useful for understanding the non-Abelian edge theories
which are relevant for the Hall effect in two-layer samples 
(e.g. the Pfaffian state) and with spinful electrons (e.g. the
Haldane-Rezayi state) \cite{mr}\cite{nonab}. 
Actually, the annulus partition functions for these edge theories have already
been proposed in Ref.\cite{read}, although their modular invariance
has not been verified explicitly. Assuming that this holds, we can 
read the value of the topological order (\ref{topan}) from the number 
of terms in the expansion of the partition function into characters: 
this is $(3m)$ for the Pfaffian theory at $\nu=1/m$, $m=2,4,\dots$,
and $(4m)$ for the Haldane-Rezayi theory at $\nu=1/m+2$ (we consider here 
the simplest case $m=2$). 
Actually, the value $(6)$ for the Pfaffian theory matches the number of
degenerate ground states found by the numerical analysis of the spectrum on 
the torus \cite{pfaff}, in agreement with our analysis of section two.
On the other hand, the value $(8)$ for the Haldane-Rezayi theory does not 
match the degeneracy $(10)$ of the trail ground-state wave function on a torus
found in Ref.\cite{esko}; nevertheless, this is a special, non-unitary
conformal theory which requires further studies.
In conclusion, we would like to stress the importance of the partition 
function for charting the Hilbert spaces of these theories, which are more 
involved than the previous Abelian theories, as well as for exposing
their symmetries.

\noindent{\bf Acknowledgements}

Conversations with J. Fr\"ohlich, V. Kac, J. Jain, N. Read 
and C. A. Trugenberger were very helpful for the development of this work.
We also thank the hospitality
of the Benasque Center for Physics, where part of this work has been done.
G.R.Z. would like to thank I.N.F.N. (Italy) and the Physics Departments of
the Universities of Torino and Firenze for their support and hospitality.
A.C. was supported in part by the European Community program
``Human Capital and Mobility''. The work of G.R.Z. is supported in part 
by a grant of the Antorchas Foundation (Argentina).

%\vfill\eject

\appendix

%- A --------------
\appendix
\section{Modular forms and functions}

In this appendix, we briefly review some of the properties of
the modular functions of interest to us.
We consider functions on a torus, which can be realized as
a fundamental domain (or unit cell) of the quotient of the
complex plane by a lattice $\Lambda$ of translations generated by two
independent periods $w_1$ and $w_2$. Any other pair of periods
$\omega^{'}_1$ and $\omega^{'}_2$ would generate the same lattice
provided the relation between this and the original pair of
periods is both ways linear with integral coefficients and
unit determinant. The torus is characterised by a single variable
$\tau =\omega_2/\omega_1$,
after rotation and dilatation invariance are considered.
The above transformations act on $\tau$ as
\beq
\tau^{'} = \frac{a\tau +b}{c\tau +d}\ , \qquad ad-bc = 1\ .
\label{fracl}
\eeq
Note that changing the sign to every coefficient yields the same
transformation. The group of these transformations is known
as the {\it modular group} $\Gamma \equiv PSL(2,\Z )= SL(2,\Z )/\Z_2$,
which is generated by two transformations $T\ :\ \tau\to\tau + 1$
and $S\ :\ \tau\to -1/\tau$, satisfying the relations
$S^2=\left(ST\right)^3 =1$ \cite{knopp}. Moreover, by a modular
transformation it is always possible to take $\tau$ to
belong to the fundamental domain
\beq
{\cal F} = \left\{ -\frac{1}{2} < \R \tau \le \frac{1}{2}\ ,
\ \vert\tau\vert^2 \ge 1\ (\R \tau \ge 0)\ ,\
\vert\tau\vert^2 > 1\ (\R \tau < 0)\ \right\}\ .
\label{fundr}
\eeq
The fermionic excitations present in the quantum Hall effect only
allow invariance under the $S$ and $T^2$ transformations.
They generate the subgroup $\Gamma_\theta \subset \Gamma$ which is
isomorphic to $\Gamma^0(2)=T\Gamma_\theta T^{-1}=\left\{
(a,b,c,d) \in \Gamma \vert b=0 \ {\rm mod}\ 2 \right\}$.
A smaller, normal, subgroup is $\Gamma (2)= \left\{
(a,b,c,d) \in \Gamma \vert a=c=1 , b=d=0 \ {\rm mod}\ 2 \right\}$,
which is generated by $T^2$ and $ST^2S$.

In the study of the functions on the torus, one naturally
encounters the modular forms $F(z)$, which transform under (\ref{fracl}) as
\beq
F\left( \frac{a\tau +b}{c\tau +d}\ \right) = \varepsilon\ 
\left(c\tau +d\right)^{\beta} F(\tau)\ ,
\label{modf}\eeq
where $\varepsilon$ is a phase and $\beta$ is the weight of the modular form.
A modular function has weight $\beta=0$.
The simplest example of a modular form is the Dedekind function,
\beq
\eta\left(\tau \right) =q^{\disp 1/24}\
\prod_{k=1}^\infty \left( 1-q^k \right)\
= q^{\disp 1/24}\ \sum_{n\in \Z} (-1)^n q^{\disp\ {1\over 2} n(3n+1)}\ ,
\qquad q={\rm e}^{\disp\ i2\pi \tau}\ .
\label{dede}\eeq
The last equality is known as Euler's pentagonal
identity, and it is a consequence of Jacobi's triple product
identity \cite{gins}:
\beq
\prod_{n=1}^{\infty} (1-q^n ) (1 + q^{n-1/2} w ) (1+q^{n-1/2}w^{-1})
= \sum_{n\in{\rm Z}} q^{n^2/2}  w^n \ ,
\label{jacobi}
\eeq
after replacing in (\ref{jacobi}) $q\to q^3$ and $w\to -q^{-1/2}$.
Under the two generators $T:\ \tau\to\tau +1$ and $S:\ \tau\to -1/\tau$
of the modular group, the transformations laws of $\eta(\tau)$ are
\barr
T:\ \eta(\tau +1) & = & {\rm e}^{\disp\ 2i\pi/24} \eta(\tau)\
,\label{tteta}\\
S:\ \eta(-1/\tau) & = & \left( -i\tau\right)^{1/2} \eta(\tau)\ .
\label{tseta}\earr
The proof of Eq.(\ref{tteta}) is straightforward, and that of
Eq.(\ref{tseta}) follows from the
application of Poisson's resummation formula
\beq
\sum_{n\in \Z} f(n)\ =\ \sum_{p\in \Z}\ \int_{-\infty}^{+\infty}
dx\ f(x)\ {\rm e}^{\disp\ 2i\pi px}
\label{poisson}\eeq
to the r.h.s. of Eq.(\ref{dede}).
Under a general transformation (\ref{fracl}) we therefore have
\beq
\eta\left(\frac{a\tau +b}{c\tau +d}\right) = \varepsilon_A
\left(c\tau +d\right)^{1/2} \eta(\tau)\ ,
\label{tetasl}
\eeq
where $\varepsilon_A$ is a 24$th$ root of unity. Thus,
the Dedekind is a modular form of weight $1/2$.

Another important example of a modular form considered
in section two is the theta function with characteristics
$a$ and $b$, which is a map ${\cal F}\times{\bf C}\to{\bf C}$ 
defined by
\beq
\Theta\left[{ a \atop b}\right]
\left(\zeta\vert\ \tau\right) = \sum_{n\in \Z}\
{\rm e}^{\disp\ i\pi \tau (n+a)^2 + i2\pi (n+a)(\zeta + b) }\ .
\label{thetaab}\eeq
The case of interest to Eq.(\ref{thetaf}) is for $a=\lambda/p$
and $b=0$, with $\lambda =1,2,\dots,p$.
The transformation properties of (\ref{thetaf}), Eq.(\ref{chitr})
follow easily from its definition. The only non-trivial
calculation regards the $S$ transformation, which can be done
following the example of the Dedekind function.
It is also easy to verify that (\ref{thetaab}) is a modular
form of weight $1/2$. It follows that the quotient
of the theta function (\ref{thetaab}) by the Dedekind function
(\ref{dede}) is a modular function.
Multi-dimensional generalisations of the results of this appendix
are straightforward, and lead to the character (\ref{mthetaf})
and transformation properties (\ref{mchitr}).

%- B -------------------------

\section{Minimal model non-invariant partition functions}

In section four we have considered the $\u1 \times\suem $ theories,
and found that their characters (\ref{thetaf}) (\ref{csuem}) give a 
unitary \rep of the modular transformations 
$T^2$, $S$, $U$ and $V$, see Eqs.(\ref{chim}),(\ref{stillv}). 
Using these characters, we have build partition functions 
which are modular invariant, see Eqs.
(\ref{dsuem}), (\ref{extinv}) and (\ref{twistinv}). 
Moreover, we have shown that the characters of the degenerate $\winf$ theories
do not carry a \rep of the $S$ transformation, unless they
are summed with multiplicities grater than one, which actually build up
the corresponding $\u1 \times\suem $ characters.
Therefore, the $\winf$ minimal models, defined in \cite{ctz5} as a 
collection of $\winf$ \reps, each counted only once, 
do not possess a modular invariant partition function
of the standard type of rational conformal field theories (\ref{rcftz}).
We do not presently know whether an alternative, non-rational 
partition function can be defined.

In this appendix, we report a simple exercise which shows that the
$S$ invariance is rather important and cannot simply be removed from the
set of self-consistency building criteria.
We relax the $S$ condition and find that many partition functions of $\winf$ 
minimal models satisfy the remaining conditions $(T^2,U,V)$, the 
closure of the fusion rules and the electron conditions. 
The minimal $\winf$ characters are of the form
$\chi^{\u1}_{\lambda}\ \chi^{W}_{\alpha}$, where the $\u1$ part
is standard (Eq.(\ref{thetaf})) and $\chi^{W}_{\alpha}$ sums one copy of all 
the ${\cal W}_m$ \reps with $SU(m)$ weight of given $m$-ality 
$\alpha=1,\dots,m$ \cite{fz}.
The $m$-ality is additive mod $m$, thus the fusion rules are closed
for theories having all values of $\alpha$, or a subset
$\alpha=\delta, 2\delta,\dots,m/\delta$, where $\delta \vert m$.
For simplicity, we consider left-right diagonal partition functions,
of the type (\ref{dsuem}), for which the $U$ condition is satisfied.
Thus, these are determined once their chiral spectrum is found;
in general, this is of the form (\ref{usspec}):
\barr
Q&=&{ml +\alpha \over \hat p}\ ,\qquad\qquad \nu= {m \over \hat p} \ ,\nl
J&=&{\left( ml +\alpha \right)^2 \over 2m\hat p}+{\alpha(m-\alpha)\over 2m}
+r\ ,\quad r\in {\bf Z}\ ,
\label{uspec2}\earr
with $\hat p$ free.
The value of $\nu$ is determined by the $V$ condition as shown later.

Starting from Eq.(\ref{uspec2}), one imposes the existence
of an excitation with the quantum numbers of the electron,
namely $Q=1$ and $2J$ an odd integer, which leads to
\barr
m{\hat l} +{\hat\a} &=& {\hat p}\ ,\nl
{\hat l} + {\hat\a} -\frac{\hat{\a}(\hat\a -1)}{m} &=& 1\ {\rm mod}\ 2 \ ,
\label{elconb}
\earr
where $\hat\a$ and $\hat l$ are the labels of the excitation
corresponding to the electron.
The condition of locality (integrality of the relative statistics)
between an arbitrary excitation in each sector $\a$ and the
electron in the $\hat\a$ sector is (see Eq.(\ref{estat})):
\beq
\a({\hat \a}-1) = 0\ {\rm mod}\ m\ .
\label{relcon}
\eeq
Moreover, the $T^2$ condition is satisfied by the solutions of 
(\ref{elconb}),(\ref{relcon}).
Note that the full spectrum of $\winf$ \reps $\a = 1,\dots ,m$ requires 
$\hat\a =1$. As explained in section 4.2, this is also the condition that
arises upon imposing the $S$ representation.
However, there are other solutions for $\hat\a \neq 1$, which 
involve the subsectors  $\alpha=a\delta$, with $\delta \vert m$.
These additional solutions were not considered in our earlier
discussion of the $\winf$ minimal models \cite{ctz5}.

Consider first the general solution to \eq{elconb}, which can be rewritten
\barr
{\hat\a}({\hat\a}-1) &=& 0\ , \ {\rm mod}\ m \nl
{\hat l} &=& \frac{{\hat\a}({\hat\a}-1)}{m} - {\hat\a}\ {\rm mod}\ 1 \ ,
\qquad \hat p= m\hat l +\hat \a \ .
\label{eits}
\earr
The solutions of ${\hat\a}({\hat\a}-1)=0\ {\rm mod}\ m$
can be found by considering each divisor $\delta$ of $m=\delta\rho$:
\beq
{\hat\a} = a\delta\ ,\qquad {\hat\a}-1 = b\frac{m}{\delta}\quad \to
\quad a\delta - b \frac{m}{\delta} =1
\label{delab}
\eeq
A solution $(a,b)$ is found whenever $\delta$ and $\rho$ are
coprime integers. 
The corresponding $\winf$ theory contains the subset of representations
$\alpha=n\delta$, $n=1,\dots,\rho$, which are solution to Eq.(\ref{relcon}) 
and are, therefore, closed under the fusion rules. 
Their spectrum is given by (\ref{uspec2}) for $\hat p=m\hat l +\hat\a$,
where $\hat l=ab -a\delta$ mod $1$.
Clearly, a common factor $\delta$ cancels in the formula for $Q$ and $\nu$;
the $V$ condition is satisfied as in Eq.(\ref{stillv}). 

In conclusion, there is a solution for any factorization of 
$m=\delta\cdot m/\delta$ with $(\delta,m/\delta)=1$.
The following solutions exist for any $m$ :
\begin{itemize}
\item 
$\delta =1$, with solution $a=1$ and
$b=0$, so that ${\hat\a}=1$ (this coincides with the Jain series).
\item
$\delta=m$, with solution $a=1$ and $b=m-1$, so that ${\hat\a}=0$
(this gives a Laughlin decoupled fluid).
\end{itemize}
The simplest solutions with non-trivial factorization of
$m$, exist for $m=6$ with $\delta=3$
and $\delta=2$. In the first case, $\hat\a =3$ and one has a
consistent theory with $SU(6)$ sectors $\a =0,3$
and filling fraction $\nu=2/(2s+3)$, $s$ even.
In the second case, $\hat\a =4$ and the sectors in the theory are
$\a =0,2,4$ and the filling fraction is $\nu=3/(3s+5)$, $s$ even.
The number of $(\hat\alpha\neq 1)$ solutions
to Eq.(\ref{delab}) grows rapidly with increasing $m$, and they
yield almost any filling fraction $\nu=n/d$.
This feature makes these $S$-variant partition functions 
inconsistent with the phenomenological pattern.

\def\NP{{\it Nucl. Phys.\ }}
\def\PRL{{\it Phys. Rev. Lett.\ }}
\def\PL{{\it Phys. Lett.\ }}
\def\PR{{\it Phys. Rev.\ }}
\def\CMP{{\it Comm. Math. Phys.\ }}
\def\IJMP{{\it Int. J. Mod. Phys.\ }}
\def\MPL{{\it Mod. Phys. Lett.\ }}
\def\RMP{{\it Rev. Mod. Phys.\ }}
\def\AP{{\it Ann. Phys. (NY)\ }}

\end{document}